\documentclass{article}
\usepackage{titling}
\usepackage{authblk}
\usepackage[english]{babel}

\usepackage[a4paper,top=2cm,bottom=2cm,left=3cm,right=3cm,marginparwidth=1.75cm]{geometry}

\usepackage{amsmath}
\usepackage{graphicx}
\usepackage[colorlinks=true, allcolors=blue]{hyperref}
\usepackage[font={small,it}]{caption}
\usepackage{subcaption}
\usepackage[labelformat=simple]{subcaption}

\usepackage[export]{adjustbox}
\usepackage{bbm}
\usepackage{comment} 
\usepackage{appendix}
\usepackage{amsmath,amssymb}
\usepackage{cancel}
\usepackage[table]{xcolor}
\usepackage{diagbox}
\usepackage{natbib}
\usepackage{mathtools}
\mathtoolsset{showonlyrefs=true}
\setcitestyle{authoryear,aysep{()}}
\usepackage{booktabs}

\graphicspath{{figures}}

\title{A Dirichlet stochastic block model for composition-weighted networks}

\author[1,2]{Iuliia Promskaia}
\author[1,2]{Adrian O'Hagan}
\author[2]{Michael Fop}
\affil[1]{Insight Research Ireland Centre for Data Analytics, University College Dublin, Dublin, Ireland}
\affil[2]{School of Mathematics and Statistics, University College Dublin, Dublin, Ireland}
\date{Corresponding author: \href{mailto: michael.fop@ucd.ie}{michael.fop@ucd.ie}} 
\begin{document}
%\begin{titlingpage}
\maketitle
             
%% Abstract
\begin{abstract}
Network data are prevalent in applications where individual entities interact with each other, and often these interactions have associated weights representing the strength of association. Clustering such weighted network data is a common task, which involves identifying groups of nodes that display similarities in the way they interact. However, traditional clustering methods typically use edge weights in their raw form, overlooking that the observed weights are influenced by the nodes' capacities to distribute weights along the edges. This can lead to clustering results that primarily reflect nodes' total weight capacities rather than the specific interactions between them. One way to address this issue is to analyse the strengths of connections in relative rather than absolute terms, by transforming the relational weights into a compositional format. This approach expresses each edge weight as a proportion of the sending or receiving weight capacity of the respective node. To cluster these data, a Dirichlet stochastic block model tailored for composition-weighted networks is proposed. The model relies on direct modelling of compositional weight vectors using a Dirichlet mixture, where parameters are determined by the cluster labels of sender and receiver nodes. Inference is implemented via an extension of the classification expectation-maximisation algorithm, expressing the complete data likelihood of each node as a function of fixed cluster labels of the remaining nodes. A model selection criterion is derived to determine the optimal number of clusters. The proposed approach is validated through simulation studies, and its practical utility is illustrated on two real-world networks.
\end{abstract}

%% Keywords

\vspace{5mm}

\section{Introduction} \label{Introduction}

Networks are often used to represent relationships between entities, and in the recent past there has been a rapid increase in the collection and availability of such relational data. Network data appear in a diverse range of applications, such as the social sciences \citep{sbm_interaction_lengths}, biology \citep{Daudin_2011}, transportation \citep{weighted_sbm}, and finance \citep{Financial_networks}. Given the wide availability of network data, the statistical analysis and modelling of networks is an active area of research, driven by the unique challenges that these data present, largely due to the dependence between entities. For a concise overview of the main models and methods for network data see \cite{network_review}.

One common task in network analysis is clustering, which involves finding groups of nodes that share similar connectivity patterns or have similar characteristics. The two main families of approaches to clustering in networks are algorithm-based and model-based \citep{clustering_in_networks_review}. Algorithm-based methods, as the name suggests, involve a set of operations that partition the network into clusters. These methods are often based on some notion of distance or similarity between nodes \citep{communities_random_walks}, spectral properties of the Laplacian matrix \citep{spec_clustering}, or an optimisation procedure \citep{guimera_2010,clauset_2005}. Model-based methods assume the presence of sub-populations in the data, corresponding to the clusters, which are then modelled by means of probability distributions. As opposed to algorithm-based methods, model-based clustering approaches have the advantage of providing a principled framework where the clustering task is framed as an inferential task for a generative probability model \citep{Daudin_2011,bouveyron_2019}.

Stochastic block modelling is a popular model-based clustering framework for network data, which models the connectivity patterns between pairs of nodes, assuming their stochastic equivalence: the presence and the strength of a connection between two nodes is determined by their cluster membership rather than by the nodes themselves. The definition of a stochastic block model as a way to describe the a priori community structure in a network was introduced in \cite{Holland83} and \cite{WangWong87}, and the first attempt to use this model as a clustering tool was made in \cite{Wasserman87}.

The stochastic block model (SBM) was originally developed for the analysis and clustering of binary network data. For this type of data, binary interactions are often defined on a pre-specified threshold or on a subjective evaluation of what constitutes an interaction between two nodes, resulting in a potential information loss. With network data becoming more widespread and complex in the recent years, a number of extensions of the SBM to networks with different edge types and characteristics have been proposed. Particular efforts have been focused on clustering models for weighted networks. For instance, \cite{Nowichi2001} extended the SBM to allow for finite set valued edges, such as those taking values in $\{-1,0,1\}$. Count edge attributes or multiple edges between pairs of nodes were considered in \cite{binomial_sbm}, \cite{Karrer-Newman2011}, \cite{Zanghi_2010}, and \cite{Mariadassou_2010}. To cluster nodes in networks with continuous edge weights, \cite{weighted_sbm} propose a weighted SBM based on gamma distributions. More general frameworks encompassing a wider range of edge types and attribute distributions have been proposed by \cite{Aicher_2013,Aicher_weighted_sbm}, and \cite{Ludkin_2020}. We point the reader to \cite{SBM_review} for a comprehensive review of extensions of the SBM for different data types as well as inferential approaches. An approach related to weighted SBMs is the one introduced by \cite{Melnykov_edge_clustering}. In \cite{Melnykov_edge_clustering}, the authors propose a general framework for performing model-based clustering on multi-layer networks. The approach uses a multivariate Gaussian distribution to model the weights, allowing the correlation between weights connecting pairs of nodes as well as between edge weights across different layers to be accounted for. This work can be seen as a generalisation of a weighted SBM \citep{Aicher_2013,Aicher_weighted_sbm,weighted_sbm} that takes into account the correlation structure between the weights.

All of the aforementioned approaches propose models for the analysis of the weights in their original form. However, in some applications, in particular those modelling flows in the network (such as transportation or trade), the observed edge weights are influenced by the nodes' weight capacities to send and receive weights along the edges, which can differ significantly across the network. This often results in clustering solutions that mimic the patterns related to the total weight capacities of the sending and receiving nodes, rather than the individual weights connecting them. Note that by the term {\em weight capacity} we mean the total volume of the weights each node can send or receive along the edges, which differs from both the total number of edges a node can send or receive and its degree \citep[see for example][]{Karrer-Newman2011}.
Taking as an example the Erasmus programme network presented in Section \ref{erasmus_application}, suppose we are interested in grouping European countries based on similarities in students' preferences for exchange destinations. With the participating countries having very different populations, we expect larger countries, such as Germany, France and the UK, to have larger student populations and hence larger volumes of students participating in the programme. These countries are also able to accommodate more incoming students. By contrast, the Nordic countries, like Denmark and Finland, are much smaller in population, and hence they send and receive significantly fewer students. If we were to perform clustering using the raw counts of students going on Erasmus exchange between these countries without taking the differences in population into account, the countries would be roughly divided into groups based on their population sizes, i.e. Germany, France and the UK as one cluster, and Denmark and Finland as another, due to differences in the scale of the weights. Even if the preference profile of the UK students is closer to those of the smaller Nordic countries, the volume of students the country sends is far too large for it to be assigned to the same cluster as Finland and Denmark. Similar issue can arise in flight networks, with small airports only being able to service a limited number of flights per day as opposed to international hubs with multiple runways, or international trade networks, where the value of goods traded is linked to country's GDP \citep{Melnykov_edge_clustering}. The differences in the magnitudes of edge weights can significantly affect the results, making it more challenging to draw conclusions. 

One way to address this is by using relative rather than absolute weights in the network. In the aforementioned cases, this can be done by calculating proportions of flow alongside each edge with respect to a sending (or receiving) node, leading to edge weights of relative size. In other applications, the absolute weights can simply be unavailable for the analysis as the interactions between nodes are measured in relative terms to begin with. For example, in epidemiological networks the edge weights could represent probabilities of interaction between agents, or in infrastructure networks they could correspond to proportions of liquid flowing between units. Networks with relative edge weights can also arise as a result of privacy protection policies, such as in \cite{Financial_networks}, where the absolute information could potentially be used to identify individual actors, banks in this case. 

The main challenge associated with the relative nature of edge weights in the network is their interdependence. Proportional data impose a constant-sum constraint, making the standard tools used for networks with independent weights invalid. To appropriately address this constraint, tools from compositional data analysis can be employed. Compositional data describe parts of some whole, such as percentages, probabilities, and proportions, and they can be naturally observed or constructed from other types of data, such as continuous or count data, by dividing individual components by their sum. There are two main families of approaches to compositional data analysis. The first involves mapping the compositions onto the real line, allowing one to use the standard machinery on the transformed data. Log-ratio transformations are a popular choice in the literature due to their sound theoretical properties, and other alternatives, such as standardisation, are also sometimes used in practice \citep{Aitchison1986, Baxter1995}. The second approach is based on direct modelling of compositional observations, which often uses the Dirichlet distribution to model the compositions in their original form \citep{Dir_clustering,dir_reg_with_zeros}. Both approaches have their relative benefits, such as the scale-invariance of the log-ratio transformations, or desirable distributional properties that come with direct modelling, therefore, the choice of the most suitable approach is context dependent. For a discussion of the main methods and challenges in compositional data analysis, we suggest the review paper by \cite{Alenazi_coda_review}. 

The literature on models and methods for networks with weights of relative nature is very scarce, with most works addressing unique application-specific challenges, often related to biological networks. Examples of such works are \cite{bioinformatics_coda_network} and \cite{bioinformatics_zero_coda_network}, where the centered log-ratio transformation is applied to the compositional data to eliminate the constant-sum constraint in an attempt to recover the microbiome interaction network. Another notable work by \cite{Financial_networks}, which is driven by a financial market application, uses direct modelling of proportions via the Dirichlet distribution and proposes a dynamic latent variable modelling framework for the analysis of interbank networks.

To the best of our knowledge, there have been no attempts to develop a general model-based clustering framework for composition-weighted networks. In this work, we propose an extension of the stochastic block model for dense networks of moderate size (in terms of the number of nodes) that utilises the relative rather than absolute strength of connections, by leveraging ideas from compositional data analysis. We model the set of edge weights from each node using a Dirichlet distribution, the parameters of which are determined by the cluster labels of the sender and each of the receivers. For inference, we consider the hybrid maximum likelihood approach developed by \cite{HybridML}, using a variant of the classification expectation-maximisation (EM) algorithm \citep{ClassificationEM}. To address model selection, an integrated completed likelihood (ICL) criterion is derived \citep{Daudin_mm_random_graphs}. We also briefly introduce an alternative approach to cluster nodes in composition-weighted networks by making use of the log-ratio transformation that maps compositional data onto the real line \citep{Aitchison1982}. The code implementing the modelling framework in R \citep{r_reference} is available on \href{https://github.com/iuliiapromskaia/dirSBM}{GitHub}. 

The structure of this paper is as follows: we start by describing the SBM with compositional Dirichlet-distributed edge weights in Section \ref{Dirichlet_sbm}, and outline the inferential procedure based on a classification EM algorithm in Section \ref{inference}; Section \ref{alternative_approach} presents an alternative method to cluster nodes in the network with compositional edge weights by using the log-ratio transformation of the compositions and then utilising a standard weighted stochastic block model; we test the model performance on synthetic data in Section \ref{sim_studies}, assessing different initialisation strategies, parameter estimation quality, and clustering and model selection performance, also in comparison to alternative approaches; we then illustrate the use of the proposed model in application to the Erasmus programme data from \cite{Erasmus_data} and the London bike sharing data from \cite{TFL} in Section \ref{real_data}; Section \ref{discussion} concludes the paper with a discussion, highlighting the work's impact and the potential for further developments.

\section{Dirichlet stochastic block model} \label{Dirichlet_sbm}
This section introduces the Dirichlet stochastic block model (DirSBM) for networks with compositional edge weights. Let $\mathbf{Y}$ be the weighted adjacency matrix of a directed network on $n$ nodes with no self-loops with strictly positive entries, i.e. $y_{ij}>0$ for all $i \neq j$ and $y_{ii}=0$ for all $i$. Define the compositional counterpart of $\mathbf{Y}$, denoted $\mathbf{X}$, whose row entries $\mathbf{x}_i$ are given by

\begin{equation}
    \mathbf{x}_i = \bigg( \frac{y_{i1}}{\sum_{j\neq i}^{n}y_{ij}},  \frac{y_{i2}}{\sum_{j\neq i}^{n}y_{ij}},\ldots,\frac{y_{i(i-1)}}{\sum_{j\neq i}^{n}y_{ij}},0,\frac{y_{i(i+1)}}{\sum_{j\neq i}^{n}y_{ij}},\ldots,\frac{y_{in}}{\sum_{j\neq i}^{n}y_{ij}} \bigg),
\end{equation}

\noindent i.e. $x_{ij}>0$ for $i\neq j$ and $\sum_{l} x_{il}=1 ~\forall i$.  Note that $x_{ii}=0$ due to the absence of self-loops in the original weighted network. Denote with $\mathbf{x}_i^*$ a subvector of $\mathbf{x}_i$ that excludes the zero entry $x_{ii}$, and with $\mathbf{X}^*$ the matrix formed by row vectors $\mathbf{x}_i^*, i=1,\ldots,n$. In practice, in some applications, the weighted adjacency matrix $\mathbf{Y}$ may be unavailable, with the compositional matrix $\mathbf{X}$ being the only piece of data provided. The Dirichlet stochastic block model with $K$ blocks assumes the following generative process for a network with compositional edge weights $\mathbf{X}^*$:

\begin{enumerate}
    \item Given a $K$-dimensional vector of cluster membership proportions $\boldsymbol{\theta}$, generate binary cluster allocations $\mathbf{z}_i \sim Multinom(1,\boldsymbol{\theta})$, for $i=1,\ldots,n$. The entries of these vectors, i.e. $z_{ik}=1$ when node $i$ is assigned to cluster $C_k$ and 0 otherwise. Denote with $\mathbf{Z}$ an $n\times K$ matrix of binary cluster allocations of all nodes in the network, and with $\mathbf{Z}_{-i}$ the $(n-1)\times K$-dimensional matrix of cluster allocations of all nodes in the network that are not $i$. 
    \item Let $\mathbf{A}=\{\alpha_{kh}\}_{k,h=1}^K$ be a $K \times K$ matrix with strictly positive values, corresponding to the Dirichlet distribution concentration parameters. Given the cluster allocations, generate $(n-1)$-dimensional compositional observations as 
    $$
    \mathbf{x}_{i}^*\lvert\mathbf{z}_i, \mathbf{Z}_{-i} \sim Dir(\mathbf{z}_i \mathbf{A} {\mathbf{Z}}^{\top}_{-i}),
    $$
    for $i=1,\ldots,n$.
\end{enumerate}
To aid understanding, consider the vector $\mathbf{c}$ denoting the cluster labels of the nodes in the network, such that $c_i = k$ if $i$ is a member of cluster $C_k$. The above distribution can also be written as 

\begin{equation} \label{dir_c}
    \mathbf{x}_{i}^* \lvert \mathbf{c} \sim Dir(\alpha_{c_i c_1},\ldots,\alpha_{c_i c_{i-1}},\alpha_{c_i c_{i+1}},\ldots,\alpha_{c_i c_n}).
\end{equation}
The intuition behind this formulation is as follows: for any sender $i$, the distribution of weights of the edges connecting it to other vertices depends on the cluster memberships of $i$ itself and of all of the receiver nodes $j \neq i$. If the sender $i$ and the receiver $j$ are members of the same cluster $C_k$, the corresponding parameter of the Dirichlet distribution is $\alpha_{kk}$, and this parameter value is shared for all such receivers $j \in C_k$. If instead the receiver node is in a different cluster, $j \in C_h$, the corresponding parameter is $\alpha_{kh}$. Since the model does not include any node-specific parameters, node-level heterogeneity is not accounted for in this framework.

The proposed model aims to describe the clustering structure of directed fully-connected networks, with the edge weights being expressed as proportions sent by the nodes (i.e. the sum of outgoing weights for each node is one). Note that the decision to use sending proportions rather than receiving proportions (i.e. where the sum of incoming weights for each node is one) is arbitrary. In fact, the modelling of receiving proportions can be achieved by working with the transpose of $\mathbf{Y}$ instead of $\mathbf{Y}$ itself.

Although DirSBM represents a novel approach for model-based clustering of proportion-weighted networks, it is not the first model that aims to account for the node capacities in the network. For example, in \cite{Karrer-Newman2011}, the authors propose the degree-corrected stochastic block model (DC-SBM), which allows better description of real-world networks with degree heterogeneity by introducing node-specific parameters. DirSBM and DC-SBM, however, serve different modeling purposes and consider node capacities differently. DC-SBM accounts for variations in the nodes' capacities to connect to other nodes in the network, allowing nodes assigned to the same cluster to have distinct degree distributions. DirSBM, on the other hand, does not aim to address the node degree heterogeneity. Instead, it focuses on discounting the effects of nodes' weight capacities, that is, the total weights each node can distribute along its edges, which can vary significantly across the network. DirSBM achieves this by representing relational data in a compositional format, thereby neutralising the impact of the total weight a node sends to (or receives from) others. This approach allows clustering to be informed by relative proportions. Moreover, DirSBM is also designed for fully-connected or highly dense networks whose degree distributions do not typically follow a power-law.

In addition to modeling proportional rather than absolute weights, DirSBM is also distinguished from existing SBMs (particularly weighted SBMs) by its relaxation of the independence assumption between edges in the network, given the cluster labels of their end nodes. For instance, \cite{Aicher_weighted_sbm} introduces a general weighted stochastic block model that balances the importance of edge presence and edge weights, with the latter drawn from an exponential family of distributions. However, this model assumes conditional independence: both binary edges and edge weights are independent, given the cluster labels, and edge weights are independent of the presence of binary edges. 
Similarly, \cite{weighted_sbm} proposes a hierarchical model in which gamma-distributed weights are attributed only to existing binary edges. However, like \cite{Aicher_weighted_sbm}, it assumes that binary edges and edge weights are conditionally independent, conditional on the cluster labels of their end nodes. To the best of our knowledge, DirSBM is a novel approach in its relaxation of the independence assumption between the edges.

\subsection{Properties of DirSBM}

The model statement in Section \ref{Dirichlet_sbm} implies that the distribution of weight vectors conditional on cluster labels in the whole network is shared among the nodes assigned to the same cluster, up to a permutation of the ordering of the nodes, which is assumed to be invariant for simplicity. To illustrate this, consider, for example, the distribution of compositional weights of node $i\in C_1$ and rearrange the order of the dimensions so that they are sorted by cluster assignment index in ascending order for ease of notation. Every vector of compositional weights of node $i$ belonging to cluster $C_1$, denoted $\mathbf{x}^*_{i \in C_1}$, conditional on the cluster assignments of all the remaining $(n-1)$ nodes, is distributed as

\begin{equation} \label{eg_x1}
    \mathbf{x}^*_{i \in C_1}|\mathbf{Z} \sim Dir(\underbrace{\alpha_{11},\ldots,\alpha_{11}}_{\text{$(n_1-1)$ times}},\, \underbrace{\alpha_{12},\ldots,\alpha_{12}}_{\text{$n_2$ times}},\,\ldots,\underbrace{\alpha_{1h},\ldots,\alpha_{1h}}_{\text{$n_h$ times}},\,\ldots,\underbrace{\alpha_{1K},\ldots,\alpha_{1K}}_{\text{$n_K$ times}}),
\end{equation}

\noindent where $n_h$ denotes the membership count of cluster $C_h$. This is because any member of cluster $C_1$ is connected to the remaining $(n_1-1)$ members of its own cluster, all $n_2$ members of cluster $C_2$, and so on. 

Generalising for a generic node in cluster $C_k$, $k = 1, \ldots, K$, we obtain:
\begin{equation} \label{eg_xk}
    \mathbf{x}^*_{i \in C_k}|\mathbf{Z} \sim Dir(\underbrace{\alpha_{k1},\ldots,\alpha_{k1}}_{\text{$n_k$ times}},\, \ldots,\underbrace{\alpha_{kk},\ldots,\alpha_{kk}}_{\text{$(n_k - 1)$ times}},\,
\ldots,\underbrace{\alpha_{kh},\ldots,\alpha_{kh}}_{\text{$n_h$ times}},\,\ldots,\underbrace{\alpha_{kK},\ldots,\alpha_{kK}}_{\text{$n_K$ times}}).
\end{equation}
Therefore, all the nodes belonging to the same cluster are identically distributed and share the same Dirichlet distribution, whose parameter vector is determined by the cluster allocations of the remaining nodes. This property of the DirSBM can be seen as a counterpart of stochastic equivalence in other stochastic block models \citep{SBM_review,Nowichi2001,snijders:1997}. The nodes in the network are called stochastically equivalent if they have the same probability of being connected by an edge to any other node in the network, conditional on their cluster labels. In weighted SBMs, such as the ones proposed by \cite{Aicher_weighted_sbm} and \cite{weighted_sbm}, this leads to identical distributions of edge weights for any pair of nodes $i\in C_k$ and $j\in C_h$ as the parameters of such distributions are only determined by the cluster labels of nodes. The DirSBM cannot be said to strictly exhibit stochastic equivalence as it does not model the probability of edge existence between nodes, but it does share the benefits that arise as a consequence of stochastic equivalence as the identical distribution property of a set of edge weights conditional on cluster labels still holds.

\subsection{Interpretation of parameters} \label{interpretation_of_params}

The parameters of the Dirichlet distribution in their original form are usually difficult to interpret as they are not measured on the same scale as the compositional data. To relate the parameter values to the original compositions, we can compute the expected proportion sent from any node $i\in C_k$ to any node $j\in C_h$, denoted $w_{kh}$. In our case, as the parameter values are shared among the pairs of nodes that belong to the same cluster pair, this expected value is given by:

\begin{equation} \label{W_matrix}
   w_{kh} = \mathbb{E}[X^*_{ij}] = \frac{\alpha_{kh}}{(n_k-1)\alpha_{kk}+\sum_{g \neq k}n_g \alpha_{kg}} \qquad i\in C_k,j \in C_h.
\end{equation}

\noindent We can use these values to construct a $K\times K$ matrix of expected exchange proportions, denoted $\mathbf{W}$, that is more intuitive to interpret than the original Dirichlet parameter matrix $\mathbf{A}$. Note that, although the entries of this matrix are proportions, the rows do not sum to 1, since these proportions are shared among all pairs of nodes with the same cluster labels.

It may also be insightful to learn the expected total shares of exchanges between clusters. To calculate these, we make use of the aggregation property of the Dirichlet distribution, which states that if the parts of the Dirichlet observation are added together, the distribution remains Dirichlet, with the updated parameter values being equal to the sum of the corresponding original values \citep{Dir_dist}. Using Equation \eqref{eg_xk}, we aggregate the parts from the same cluster pair, resulting in a compositional vector 
$$
\mathbf{q}_{i}|\mathbf{Z}=\left(\sum_{\substack{j\neq i \\ j\in C_1}} x_{ij}, \ldots, \sum_{\substack{j\neq i \\ j\in C_k}} x_{ij}, \ldots \sum_{\substack{j\neq i \\ j\in C_K}} x_{ij} \right) \sim Dir(n_1\alpha_{k1},\ldots,(n_k-1)\alpha_{kk},\ldots,n_K\alpha_{kK}),
$$
for $i\in C_k$. Hence, the expected cluster-to-cluster exchange shares, denoted with $v_{kh}$, are given by:

\begin{equation} \label{V_matrix}
   v_{kh} = \mathbb{E}[q_{il}]  =  
   \begin{cases}
       \dfrac{(n_k-1)\alpha_{kk}}{(n_k-1)\alpha_{kk}+\sum_{g \neq k}n_g \alpha_{kg}} \text{\hspace{2mm} if } h=k,\\[15pt]
       \dfrac{n_h\alpha_{kh}}{(n_k-1)\alpha_{kk}+\sum_{g \neq k}n_g \alpha_{kg}} \text{\hspace{2mm} if } h\neq k. 
       \end{cases}
\end{equation}

\noindent The resulting matrix $\mathbf{V}=\{v_{kh}\}_{k,h=1}^K$ exhibits a unit-row-sum property.

\subsection{Networks with zero-weighted edges} \label{zero-weights}
As stated earlier, DirSBM is developed for fully connected networks characterized by weighted adjacency matrices with strictly positive entries. However, as shown in the applications presented in Section \ref{real_data}, even the most dense real-world networks exhibit some percentage of zeros in the weighted adjacency matrix. These zeros in the adjacency matrix may represent either the absence of edges between certain pairs of nodes, or the presence of edges with zero weight. The distinction between the two often affects the modelling process, as the former relates to network sparsity and the latter does not. An edge is absent in the network where a connection between the nodes is not possible or has not been observed. A zero-weighted edge, on the other hand, is defined as a present connection between the nodes that is attributed with a zero weight or strength. For instance, in the context of the Erasmus network from Section \ref{erasmus_application}, exchange between any pair of countries is possible, therefore zeros in the raw data matrix can be viewed as zero weights. 

Since DirSBM is designed for clustering fully-connected networks, modelling network sparsity is outside the scope of this work. Consequently, we treat any zeros in the weighted adjacency matrix as zero-weighted edges rather than absent links. To address zero-weighted edges, we adopt a standard procedure from compositional data analysis, where zeros present unique challenges and remain an active area of research, as standard tools are not designed to handle zero-valued compositional parts \citep{coda_book}. A common approach is to replace zeros with a small constant value, as this method is simple and generally effective. Therefore, to address zero-weighted edges in the data, a small value $\epsilon$ is added to the raw data before converting it to compositions. The final results are typically robust relative to the magnitude of the added constant since the small value is added to the raw data rather than to the final compositions. In Section~\ref{zero_sim_study} we assess the impact of the proportion of zero-weighted edges in the network on parameter estimation and clustering performance of DirSBM.

\section{Inference} \label{inference}

The distribution of a set of edge weights $\mathbf{X}$, conditional on the cluster indicator variable is

\begin{equation}
    p(\mathbf{X}|\mathbf{Z}) = \prod_{i=1}^n p(\mathbf{x}_i| \mathbf{z}_i,\mathbf{Z}_{-i}) = \prod_{i=1}^{n} \prod_{k=1}^{K} \Bigg [\frac{\Gamma (\sum_{j\neq i}^{n} \alpha_j)}{\prod_{j\neq i}^{n} \Gamma (\alpha_j)} \prod_{j\neq i}^{n} x_{ij}^{\alpha_j-1} \Bigg ] ^{z_{ik}}, \text{\hspace{1mm} where } \alpha_j=\sum_{h=1}^{K} z_{jh} \alpha_{kh}.
\end{equation}
This is a product of probability densities of $(n-1)$-dimensional Dirichlet distributions with the respective parameter vectors determined by the cluster label of the sender node $i$ and of the remaining nodes in the network. 

As usual in model-based clustering, inference proceeds by maximizing the marginal log-likelihood of the data $\mathbf{X}$ with respect to the model parameters. However, this quantity involves summing over the set of all possible cluster allocations and is computationally intractable \citep{Daudin_mm_random_graphs}. A popular algorithm used in model-based clustering is the expectation-maximisation (EM) algorithm \citep{Dempster-EM}. It is employed to find the maximum of the marginal log-likelihood of the data by introducing latent variables corresponding to cluster labels and defining the joint log-likelihood of the data and the cluster labels, the complete data log-likelihood. Then, inference for the model consists of estimation of parameters and of the latent cluster assignments. The EM algorithm works by iteratively finding the expectation of the complete data log-likelihood with respect to the posterior distribution of the latent variables (E-step) and maximising this expectation with respect to the model parameters (M-step).  In the case of the DirSBM, the complete data log-likelihood is:

\begin{equation} \label{Complete_ll}
\begin{aligned}
    l_c(\mathbf{A},\boldsymbol{\theta})& = \log p(\mathbf{X},\mathbf{Z})\\& = \sum_{i=1}^{n} \sum_{k=1}^{K} z_{ik} \Bigg [ \log \Gamma (\sum_{j\neq i}^{n} \alpha_j) - \sum_{j\neq i}^{n} \log \Gamma (\alpha_j) + \sum_{j\neq i}^{n} (\alpha_j-1) \log x_{ij} \Bigg ] + \sum_{i=1}^{n} \sum_{k=1}^{K} z_{ik} \log \theta_k.
\end{aligned}
\end{equation} 

\noindent Further details are provided in \ref{l_c}.

In stochastic block modelling, the E-step is often computationally intractable due to the fact that the latent variables are not independent a posteriori. To overcome this problem, the variational EM is often employed for inference in SBMs, which is based on an approximation of the posterior distribution of the latent variables \citep{Daudin_mm_random_graphs}. 

\subsection{Hybrid log-likelihood} \label{hybrid_log_lik}

In the DirSBM, a new challenge arises when trying to employ the variational EM for inference, as it is standard in SBMs. This is due to the fact that, in order to derive the evidence lower bound \citep{Daudin_mm_random_graphs} one is required to compute the expectation:

\begin{equation}
    \mathbb{E}_{\mathbf{Z}\sim q(\mathbf{Z})} \Bigg[ \log \Gamma (\sum_{j\neq i}^{n} \sum_{h=1}^{K} z_{jh}\alpha_{kh}) - \sum_{j\neq i}^{n} \log \Gamma (\sum_{h=1}^{K} z_{jh}\alpha_{kh}) \Bigg],
\end{equation}

\noindent with respect to $q(\mathbf{Z})$, the mean field approximation of the posterior distribution of the latent cluster allocations. This expectation is intractable, because, unlike in standard SBM, even conditional on their cluster labels, the distribution of node $i$ is not independent of that of node $j$. In fact, due to the model formulation in \eqref{Complete_ll}, one needs to know the cluster assignments of all the receiving nodes contained in $\mathbf{Z}_{-i}$ to define the distribution of $\mathbf{x}_i$ given $\mathbf{z}_i$.

For this reason, we employ the inferential approach based on a hybrid log-likelihood proposed by \cite{HybridML}. The approach uses a working independence assumption, whereby each of the latent variables corresponding to the cluster label $c_i$ is seen as a function of fixed values of the remaining cluster labels $\mathbf{c}_{-i}$, allowing factorization of the likelihood over the nodes. The observed hybrid log-likelihood, using the notation from \eqref{dir_c}, is given by

\begin{equation}
    l^{hyb}(\mathbf{A},\boldsymbol{\theta}) = \sum_{i=1}^{n} \log \left( \sum_{k=1}^{K}  \theta_k p(\mathbf{x}_i | c_i=k,\mathbf{c}_{-i}=\Tilde{\mathbf{c}}_{-i}) \right),
\end{equation}
where the notation $\Tilde{\mathbf{c}}_{-i}$ is employed to indicate the fixed nature of the cluster labels of the remaining $(n-1)$ nodes in the network when considering the contribution of node $i$ to the log-likelihood.

Introducing the set of latent variables $\mathbf{Z}$, we arrive at the complete data hybrid log-likelihood:

\begin{equation} \label{Complete_hybrid_ll}
\begin{aligned}
        l^{hyb}_c(\mathbf{A},\boldsymbol{\theta}) & = \sum_{i=1}^{n} \sum_{k=1}^{K} z_{ik} \log p(\mathbf{x}_i|z_{ik}=1,\mathbf{Z}_{-i}=\widetilde{\mathbf{Z}}_{-i}) + \sum_{i=1}^{n} \sum_{k=1}^{K} z_{ik} \log \theta_k \\&
        = \sum_{i=1}^{n} \sum_{k=1}^{K} z_{ik} \Bigg [ \log \Gamma (\sum_{j\neq i}^{n} \Tilde{\alpha}_j) - \sum_{j\neq i}^{n} \log \Gamma (\Tilde{\alpha}_j) + \sum_{j\neq i}^{n} (\Tilde{\alpha}_j-1) \log x_{ij} \Bigg ] + \sum_{i=1}^{n} \sum_{k=1}^{K} z_{ik} \log \theta_k,
\end{aligned}       
\end{equation}
\noindent where $\widetilde{\mathbf{Z}}$ is a binary matrix with entries $\Tilde{z}_{ik}=\mathbbm{1}\{\Tilde{c}_i=k\}$ and $\Tilde{\alpha}_j=\sum_{h=1}^{K} \Tilde{z}_{jh} \alpha_{kh}$. With the complete data hybrid log-likelihood, a working independence assumption allows one to implement a variant of the classification EM algorithm \citep{ClassificationEM}, where the hard partition of the nodes that are not $i$ is used to compute the cluster assignment probability for $i$. The steps of the algorithm are outlined below.

\vspace{5mm}
\noindent \textbf{E-step:} compute the probability of assignment of node $i$ to cluster $k$ at iteration $t$ using

\begin{equation}\label{e-step}
    \hat{z}_{ik}^{(t)} = \widehat{\Pr}(z_{ik}=1|\mathbf{x}_i,\widetilde{\mathbf{Z}}_{-i}^{(t-1)}) \propto \hat{\theta}_k^{(t-1)} \dfrac{\Gamma(\sum_{j\neq i}^{n} \Tilde{\alpha}_j^{(t-1)})}{\prod_{j\neq i}^{n} \Gamma (\Tilde{\alpha}_j^{(t-1)})} \prod_{j\neq i}^{n} x_{ij}^{\Tilde{\alpha}_j^{(t-1)}},
\end{equation}
\noindent with $\Tilde{\alpha}_j^{(t-1)}=\sum_{h=1}^{K} \Tilde{z}_{jh}^{(t-1)} \hat{\alpha}_{kh}^{(t-1)}$. The quantity on the right hand side is normalised to fulfill the constraint $\sum_{k=1}^{K}  \hat{z}_{ik}^{(t)}=1$.

\vspace{5mm}
\noindent\textbf{C-step:} following \cite{HybridML}, a greedy approach to the C-step is adopted. That is, for each node $i=1,\ldots,n$, every label swap is considered and $i$ is assigned to the cluster label producing the highest value of the observed hybrid log-likelihood, i.e. 

\begin{equation} \label{c-step}
    c_i^{(t)} = \underset{k=1,\ldots,K}{\arg \max} \hspace{2mm} \sum_{l=1}^{n} \log \left( \sum_{h=1}^{K}  \hat{\theta}_h^{(t-1)} p(\mathbf{x}_l | c_l=k,\mathbf{c}_{-l}=\Tilde{\mathbf{c}}_{-l}) \right),
\end{equation}

\noindent then $\mathbf{c}^{(t)}$ and $\widetilde{\mathbf{Z}}^{(t)}$ are updated. The update of cluster labels happens sequentially for the nodes in the network rather than simultaneously, therefore any changes in the cluster assignments of nodes before $i$ affect $\Tilde{\mathbf{c}}_{-i}$ and hence the cluster label for $i$. 

\vspace{5mm}
\noindent\textbf{M-step:} maximise the expectation of the complete data hybrid log-likelihood

\begin{equation}
\begin{aligned}
    \mathbb{E}[l_c^{hyb}(\mathbf{A},\boldsymbol{\theta})] = &\sum_{i=1}^{n} \sum_{k=1}^{K} \hat{z}_{ik}^{(t)} \Bigg (\log \Gamma (\sum_{j\neq i}^{n} \Tilde{\alpha}_j^{(t-1)}) - \sum_{j\neq i}^{n} \log \Gamma (\Tilde{\alpha}_j^{(t-1)}) + \sum_{j\neq i}^{n} (\Tilde{\alpha}_j^{(t-1)} - 1) \log x_{ij} \Bigg ) \\& + \sum_{i=1}^{n} \sum_{k=1}^{K} \hat{z}_{ik}^{(t)} \log \hat{\theta}_k^{(t-1)}.    
\end{aligned}
\end{equation}

\noindent Closed form solution is available for the mixing proportions, which are updated using

\begin{equation} \label{m-step}
    \hat{\theta}_k^{(t)} = \frac{\sum_{i=1}^{n} \hat{z}_{ik}^{(t)}}{n}.
\end{equation}
The updates for Dirichlet parameters are not available in closed form, and the set of estimates $\{\hat{\alpha}_{kh}^{(t)}\}_{k,h=1}^{K}$ are found numerically using the R function $optim$ \citep{r_reference}. Since the parameters of the Dirichlet distribution are strictly positive, we use the L-BFGS-B optimisation routine, which allows for optimization with parameters having a bounded range \citep{Byrd_optim}. 

The steps above are iterated until the following convergence criterion is met:

\begin{equation}
    \left| \frac{l^{hyb}(\mathbf{\hat{A}}^{(t)},\boldsymbol{\hat{\theta}}^{(t)})-l^{hyb}(\mathbf{\hat{A}}^{(t-1)},\boldsymbol{\hat{\theta}}^{(t-1)})}{l^{hyb}(\mathbf{\hat{A}}^{(t)},\boldsymbol{\hat{\theta}}^{(t)})} \right| < \varepsilon.
\end{equation}
In this work, $\varepsilon=10^{-5}$ was used. More details on the algorithm can be found in \ref{CEM_appendix}.

\subsection{Identifiability of parameter ordering}
The DirSBM exhibits a minor non-idenfitiability issue related to the ordering of clusters and the ordering of parameters. In particular, once we obtain the final node partition and the parameter estimates using the algorithm described in Section \ref{hybrid_log_lik}, the ordering of rows of the estimated Dirichlet concentration matrix $\hat{\mathbf{A}}$ and of the entries of the mixing proportions vector $\hat{\boldsymbol{\theta}}$ is not guaranteed to match the ordering of clusters in the partition. To aid understanding, consider an example with 

$$
\mathbf{c}=(1,1,2,1,2),~ \mathbf{A} = \begin{pmatrix} \alpha_{11} & \alpha_{12} \\ \alpha_{21} & \alpha_{22} \end{pmatrix},~ \text{ and } ~\boldsymbol{\theta}=(\theta_1,\theta_2).
$$
For the first observation, which is placed in cluster 1, $\mathbf{x}_1^* \sim \theta_1 Dir(\alpha_{11},\alpha_{12},\alpha_{11},\alpha_{12}) + \theta_2 Dir(\alpha_{21},\alpha_{22},\alpha_{21},\alpha_{22})$. Now, consider the same partition of nodes with the same cluster order and permute the rows of the parameter matrix $\mathbf{A}$ and the entries of $\boldsymbol{\theta}$, i.e.

$$
\mathbf{c}=(1,1,2,1,2), ~ \mathbf{A}' = \begin{pmatrix} \alpha_{21} & \alpha_{22} \\ \alpha_{11} & \alpha_{12} \end{pmatrix} = \begin{pmatrix} \alpha_{11}' & \alpha_{12}' \\ \alpha_{21}' & \alpha_{22}' \end{pmatrix},~\text{ and }~\boldsymbol{\theta}'=(\theta_2,\theta_1)=(\theta_1',\theta_2').
$$
Under the permutation of the parameters $\mathbf{A}$ and $\boldsymbol{\theta}$, the first observation is distributed $\mathbf{x}_1^* \sim \theta_1' Dir(\alpha_{11}',\alpha_{12}',\alpha_{11}',\alpha_{12}') + \theta_2' Dir(\alpha_{21}',\alpha_{22}',\alpha_{21}',\alpha_{22}')$. 
Clearly:
\begin{align*}
&\theta_1' D(\mathbf{x}_1^*; \alpha_{11}',\alpha_{12}',\alpha_{11}',\alpha_{12}') + \theta_2' D(\mathbf{x}_1^*; \alpha_{21}',\alpha_{22}',\alpha_{21}',\alpha_{22}')\\
&= \theta_2 D(\mathbf{x}_1^*; \alpha_{21},\alpha_{22},\alpha_{21},\alpha_{22}) + \theta_1 D(\mathbf{x}_1^*; \alpha_{11},\alpha_{12},\alpha_{11},\alpha_{12}) 
\end{align*}
where, for ease of notation, $D(\mathbf{x}; \ldots)$ denotes the Dirichlet density evaluated at $\mathbf{x}$ with the corresponding concentration parameters. Therefore, the permuted parameters $\mathbf{A}'$ and $\boldsymbol{\theta}'$ yield the same likelihood value as the parameters in the original order.

To address this issue, we create a hard partition based on the estimated probabilities $\hat{\mathbf{Z}}$ from the E-step. Subsequently, we use the order of clusters in this hard partition as reference since the column entries of $\hat{\mathbf{Z}}$ are tied to the individual terms of the Dirichlet mixture. Lastly, we use the $matchClasses()$ function in R \citep[$e1071$ package,][]{e1071} to match the order of clusters in the E-step to that in the C-step, and use this match to reorder the rows of $\hat{\mathbf{A}}$ and entries of $\hat{\boldsymbol{\theta}}$ at the convergence of the algorithm.

\subsection{Algorithm initialisation} \label{initialisation}
As in most mixture models, extra attention should be paid to the choice of the initialisation strategy since the objective function in question is not unimodal, and the algorithm is only guaranteed to converge to a local optimum \citep{EM_convergence}. In this paper, we consider the following initialisation strategies, which are compared in the simulation study in Section \ref{sim_study_init}:

\begin{itemize}
    \item \textbf{Random:} each node is assigned to one of $K$ groups randomly with equal probability. Starting with a random partition of nodes gives us an idea of the performance of the algorithm when no information on the data is used when the initial partition is created. To reduce the risk of convergence to a poor local maximum, we run the algorithm with a number of random starting partitions and select the best based on the value of the observed hybrid log-likelihood. In the simulation study in Section \ref{sim_study_init}, we set the number of random starting partitions equal to 5 as we want to strike a balance between performance and computational costs. A larger number of random starting partitions have been tested, providing only a marginal improvement on average, but at a higher computational cost. 
    \item \textbf{K-means:} the k-means algorithm \citep{kmeans} is repeatedly used to perform clustering on the data matrix $\mathbf{X}^*$ to reduce the risk of convergence to a local maximum, and the clustering solution of the best run is used as an initial partition. This is a widely used initialisation strategy as it is computationally inexpensive. Since the k-means algorithm itself is initialised at random, we run it 50 times in order to find the initial cluster labels as part of the simulation study.
    \item \textbf{K-means on CLR-transformed data (CLR+K-means):} as our data are compositional in nature, the centered log-ratio transformation is first applied to the data matrix $\mathbf{X}^*$ \citep{Aitchison1982}, which maps the compositions to the real line, then the k-means algorithm is used in the same fashion as above to find an initial partition \citep{k-means_on_clr_data}.
    \item \textbf{Spectral clustering:} eigendecomposition of $\mathbf{X}^*$ is performed and used for dimensionality reduction, then the k-means algorithm is used on the reduced data to get a set of initial cluster labels \citep{spec_clustering}. Spectral clustering is widely used for initialisation as it can handle more complex cluster shapes than k-means itself whilst maintaining low computational costs.
    \item \textbf{Binary SBM (BinSBM):} a binary version of the network is created from $\mathbf{X}$ by assigning binary edges where the weighted edge exceeds a pre-specified threshold. After empirical investigation, we selected as a threshold the mean weight, as it tends to provide a reasonably well-connected binary network that retains clustering structure. Once such a binary network is created, a Bernoulli stochastic block model is fitted using the R package $blockmodels$ \citep{blockmodels_package} and the clustering partition is used to initialise the algorithm.
    \item \textbf{Gaussian SBM (GausSBM):} a Gaussian stochastic block model is fitted directly to $\mathbf{X}$ using the R package $blockmodels$ \citep{blockmodels_package} and the obtained clustering solution is used for initialisation.
    \end{itemize}

\subsection{Model selection} \label{model_selection}
The modelling approach detailed in Section \ref{Dirichlet_sbm} is only applicable when the number of clusters is set in advance. However, in reality, in many situations the number of clusters is unknown and needs to be inferred from the data. Therefore, a model selection criterion indicating how well models with varying numbers of clusters fit the data is required in order to choose the number of groups appropriately. 

A popular choice in stochastic block modelling is the integrated completed likelihood criterion \citep[ICL,][]{ICL,Daudin_mm_random_graphs}, typically employed for selecting the number of groups in this context; see for example \cite{sbm_interaction_lengths}, \cite{weighted_sbm}, and \cite{binomial_sbm}. In the case of DirSBM, the ICL is given by

\begin{equation}
    ICL(K) = l^{hyb}_c(\mathbf{\hat{A}},\boldsymbol{\hat{\theta}}|\mathbf{X},\mathbf{\hat{Z}},K) - \frac{1}{2} K^2\log n(n-1) -\frac{1}{2}(K-1)\log n,
\end{equation}
where $l^{hyb}_c(\mathbf{\hat{A}},\boldsymbol{\hat{\theta}}|\mathbf{X},\mathbf{\hat{Z}},K)$ is the complete data hybrid log-likelihood from Equation \eqref{Complete_hybrid_ll}, evaluated using parameter estimates at convergence of the algorithm, used in place of the complete data log-likelihood following \cite{HybridML}. We fit the DirSBM with different numbers of clusters, and compute the respective values of the criterion. The highest ICL value corresponds to the optimal number of groups. \ref{ICL} contains derivations of the ICL that follow closely the derivations of the original ICL for stochastic block models proposed in \cite{Daudin_mm_random_graphs}. The simulation study concerning model selection performance using the proposed criterion is contained in Section \ref{sim_study_K}.

\section{An alternative approach} \label{alternative_approach}

As discussed in Section \ref{Introduction}, there are two main approaches to working with compositions in compositional data analysis. One involves applying a transformation to the data that maps them to the Euclidean space, followed by the use of standard tools and models for independent data. The second is based on direct modelling of compositions \citep{Alenazi_coda_review}. The methodology proposed in Section \ref{Dirichlet_sbm} aims to model the compositions directly, but since both approaches have their relative benefits, we propose an alternative that falls under the transformation-based approach, assessing its performance in Section \ref{sim_study_clustering} for comparative purposes.

We first map the compositions onto the real line using the centered log-ratio (CLR) transformation proposed in \cite{Aitchison1982}:
\begin{equation}
    \mathbf{u}_i=CLR(\mathbf{x}_i)=  \left( \log \frac{x_{i1}}{\bar x},\ldots,0,\ldots,\log \frac{x_{in}}{\bar x}\right),
\end{equation}
\noindent where $\bar x = \left (\prod_{j\neq i}^{n} x_{ij} \right) ^\frac{1}{n-1}$ is the geometric mean of $\mathbf{x}$, then apply SBM with Gaussian weights \citep{Aicher_2013,Aicher_weighted_sbm} to the transformed data, assuming that each individual edge weight follows

\begin{equation} \label{alternative_dist}
    u_{ij}|(z_{ik}=1,z_{jh}=1) \sim N(\mu_{kh},\sigma^2).
\end{equation}

The centered log-ratio transformation has been chosen for its property of preserving the dimensionality of the observations. That is, it maps the Aitchison simplex sample space $\mathbb{S}^{n}$ to the standard Euclidean space $\mathbb{R}^{n}$, unlike many other transformations that map to $\mathbb{R}^{n-1}$ \citep{coda_review_greenacre}. In this case, each dimension of $\mathbf{u}_i$ corresponds to a transformed weight of an edge connecting node $i$ to each of the other nodes in the network, hence the reduction of dimensionality would require a choice of an edge to be dropped, leading to information loss.

The approach of fitting the Gaussain SBM on CLR-transformed data is implemented using the $blockmodels$ and $compositions$ packages in R \citep{blockmodels_package,compositions_package}, and its clustering performance is assessed alongside the DirSBM in Section \ref{sim_study_clustering}. This approach is a rather strong competitor for the DirSBM proposed in Section \ref{Dirichlet_sbm}, in particular in simpler and smaller cases, such as in networks with relatively few nodes with a small number of clusters.

There are two major drawbacks of the alternative approach in comparison to the DirSBM. First of all, the normality assumption in Equation \eqref{alternative_dist} may not be appropriate as the mapping of compositional vectors to the real line (whether by using CLR or any other transformation) only eliminates a unit-sum constraint and does not guarantee convenient distributional behaviour of the resulting transformed data. This means that, in order to utilise this approach, one needs to check the appropriateness of the transformation and the type of weighted SBM to fit to the transformed data. The DirSBM, on the other hand, is a ready-to-use model that takes into account the unique behaviour of the compositional data as it models them directly. The second disadvantage of the transformation-based approach is interpretability. The parameters of the distribution in Equation \eqref{alternative_dist} (or of any other distribution selected to model the transformed data) refer to the auxiliary Euclidean space, and it is unclear how one could use these to understand the patterns present in the original simplex space. As presented in Section \ref{interpretation_of_params}, the parameters of the DirSBM can be used to compute the expected exchange proportions between nodes as well as clusters, making the interpretation of results intuitive.

\section{Simulation studies} \label{sim_studies}

In this section, we consider various simulation studies to assess the performance of our model with respect to clustering, parameter estimation, and model selection. We also investigate the effect of the different initialisation strategies listed in Section~\ref{initialisation} on the quality of the final inferred clustering of the nodes. Additionally, we examine the effect on the model's performance of the presence of zero-weighted edges in the network. Throughout this section, where relevant, clustering performance is evaluated using the adjusted Rand index \citep[ARI,][]{Rand,adjRand}, which measures the agreement between two partitions, adjusted for random chance, by comparing the estimated clustering against the true classification of the nodes.

Artificial data are generated according to the DirSBM presented in Section \ref{Dirichlet_sbm}. Different numbers of clusters, $K=\{2,3,5\}$, and network sizes, $n=\{30,50,70,100\}$ are considered. The parameter matrices $\mathbf{A}$ are defined so that the entries have varying levels of homogeneity, resulting in indirect modifications to the degree of overlap between clusters in the compositions. The term ``level of homogeneity" is used to describe the extent to which the between-cluster weight parameters, i.e. $\{\alpha_{kh}\}_{h \neq k}$, are close to the respective within-cluster weight parameters, $\alpha_{kk}$. When the values of the between- and the within-cluster parameters are set to be more similar, the resulting compositional weights in the network become more homogeneous, thus making separating the clusters more challenging. On the other hand, the lower the homogeneity, the more separated the clusters. Further details on the parameter values used in the simulation studies can be found in \ref{sim_study_notes}.

\subsection{Choice of initialisation strategy} \label{sim_study_init}
We consider the different initialisation strategies of Section~\ref{initialisation} and evaluate the model performance with respect to cluster label recovery on simulated data. We generate 50 networks with $n=50$ nodes with 3 similarly-sized clusters and the parameter matrix $\mathbf{A}$ is chosen so that the resulting within- and between-cluster edge weights are reasonably, but not excessively, distinct:
\begin{equation}
    \mathbf{A}= \begin{pmatrix} 1.0 & 0.7 & 0.5 \\ 0.9 & 1.5 & 0.6 \\ 0.4 & 0.5 & 1.2  \end{pmatrix}.
\end{equation}

\begin{figure}[t] 
\centering
\includegraphics[width=0.6\linewidth,valign=t]{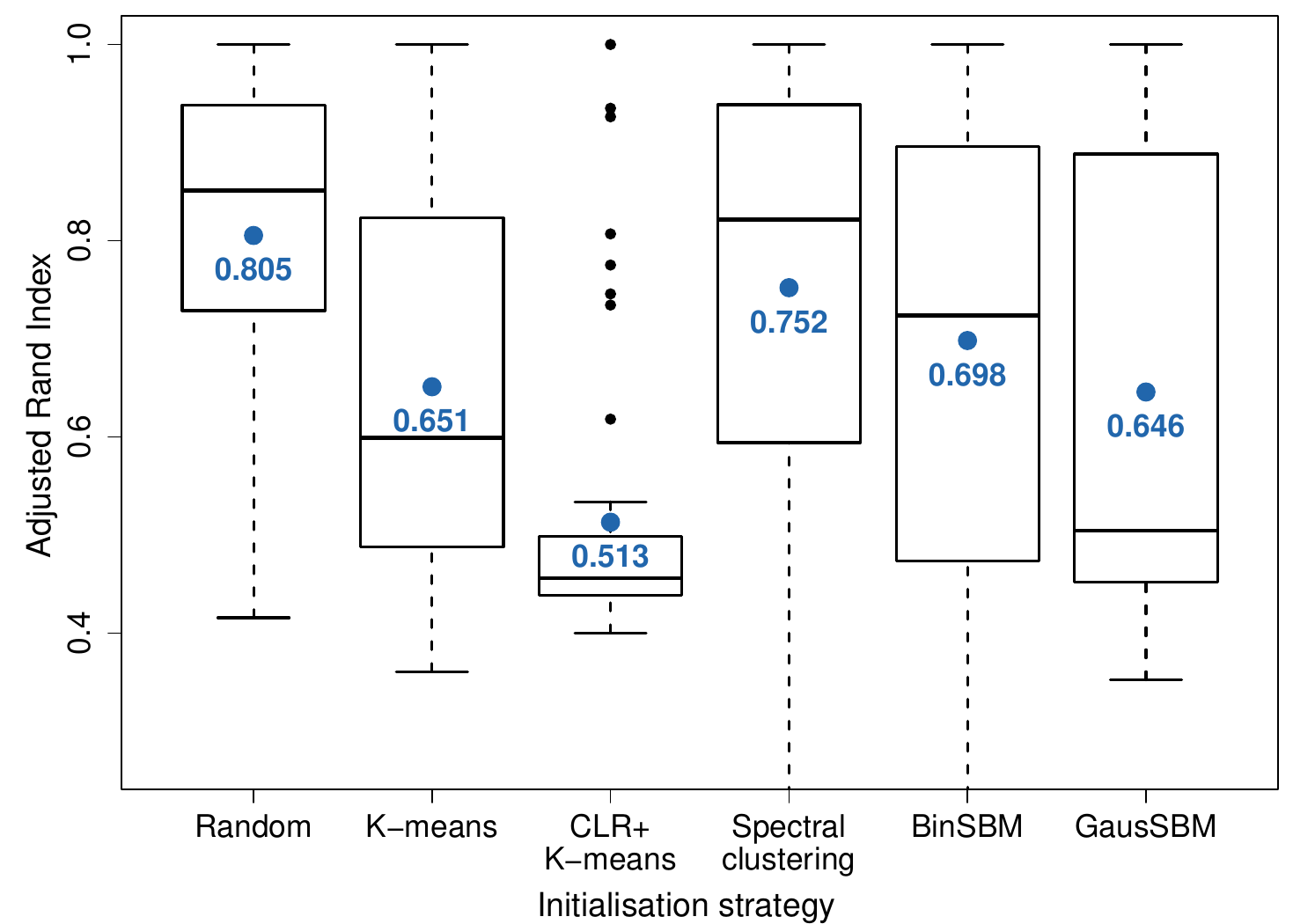}
\caption{Adjusted Rand index (ARI) of DirSBM with different initialisation strategies, with the mean value of ARI indicated in blue.}
\label{Init_boxplot}
\end{figure}

We initialise the algorithm using the strategies outlined in Section \ref{initialisation} and compute the ARI. The results of this simulation study are shown in Figure \ref{Init_boxplot}. From the figure, random initialisation seems to work best on average, with a mean ARI of $0.805$, despite being the only strategy that is not informed by the data. It also exhibits relatively low variability in comparison to all but one other strategy, which is k-means on log-ratio transformed data that is clearly inferior, with a mean ARI of only $0.513$. Further investigations also reveal that, as well as having the best average $ARI$ and low variability, random initialisation has led to the highest observed hybrid log-likelihood value in 33 out of 50 instances. Given that the number of random partitions considered was only 5, this result also shows that the proposed algorithm is capable of recovering the clustering patterns fairly successfully even when the initial cluster labels are assigned randomly and are not based on some educated guess.  Throughout the simulation studies, we use random initialisation with 5 starts, and to increase the reliability of results for the real world data in Section \ref{real_data} we increase the number of initial random partitions to 20.

It may come as a surprise to the reader that out of all the strategies considered, random initialisation performed best, as it gives a starting cluster partition that is completely uninformed. However, there are two aspects of the data modelled by the DirSBM that are worth remembering. Firstly, the data matrix $\mathbf{X}^*$ introduced in Section \ref{Dirichlet_sbm} is quite unusual in its structure in the sense that the ordering of the dimensions in the compositional edge weight vectors is not fixed and is tied to the index of the sender node. For example, the first entry of observation $\mathbf{x}^*_1$, i.e. $x^*_{11}$, refers to the weight of an edge from node 1 to node 2 (as there are no self-loops and hence there is no weight corresponding to an edge from node $1$ to itself), whereas, for any other node $i \neq 1$, $x^*_{i1}$ is the edge weight from node $i$ to node 1. The same shift along the dimensions happens across the whole set of compositional weight vectors as we take each node in turn to be the sender. This means that distance-based algorithms, such as k-means, would not necessarily compute the distances between the matching pairs of nodes' edge weights, e.g. the first set of distances for node $1$ would involve the edge weight from node 1 to node 2, and for any other node $i$, it would involve the edge weight from node $i$ to node 1. This can potentially lead to finding spurious clustering patterns, producing initial partitions that result in convergence to local maxima.

The second aspect is the fact that the edges in the networks are not independent, because the weights are defined in terms of compositions. Using initialisation strategies that assume independence between the edge weights, we risk identifying false patterns in the network and getting stuck in a local maximum as a result. For instance, the Gaussian SBM initialisation could give rise to a misleading initial partition as the unit-sum constraint is ignored. With repeated random initialisation, we are less likely to get stuck in one of the local maxima as we can control the exploration of the partition space by changing the number of random starts.

\subsection{Parameter estimation performance} \label{sim_study_params}
In this section we examine parameter estimation quality when the number of clusters is known in advance. We consider the cases with 2, 3 and 5 clusters as well as low, medium and high level of parameter homogeneity, as described at the start of Section \ref{sim_studies}. The size of the networks is also varied to inspect its effect on parameter estimation quality. 

\begin{figure}[!t] 
\centering
\begin{subfigure}{0.63\textwidth}
\includegraphics[width=1\linewidth,valign=t]{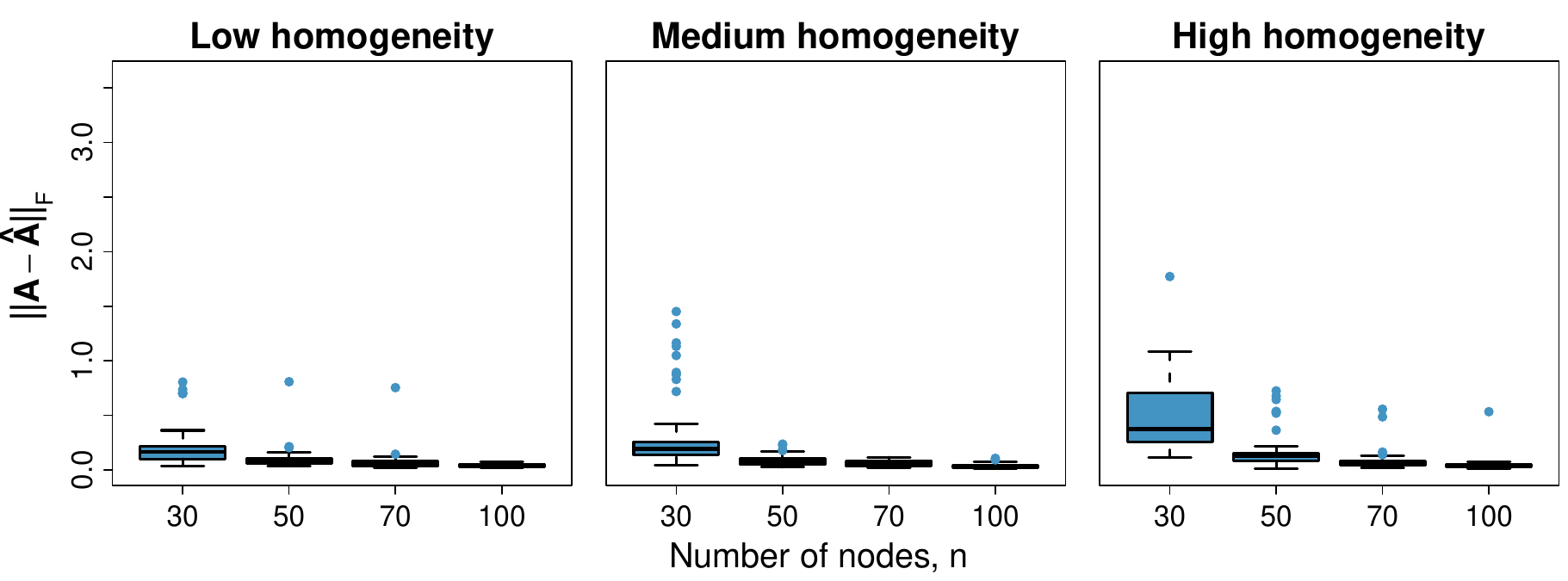}
\subcaption{K=2} \label{param_rec_K=2}
\vspace{1.5mm}
\end{subfigure}
\begin{subfigure}{0.63\textwidth}
\includegraphics[width=1\linewidth,valign=t]{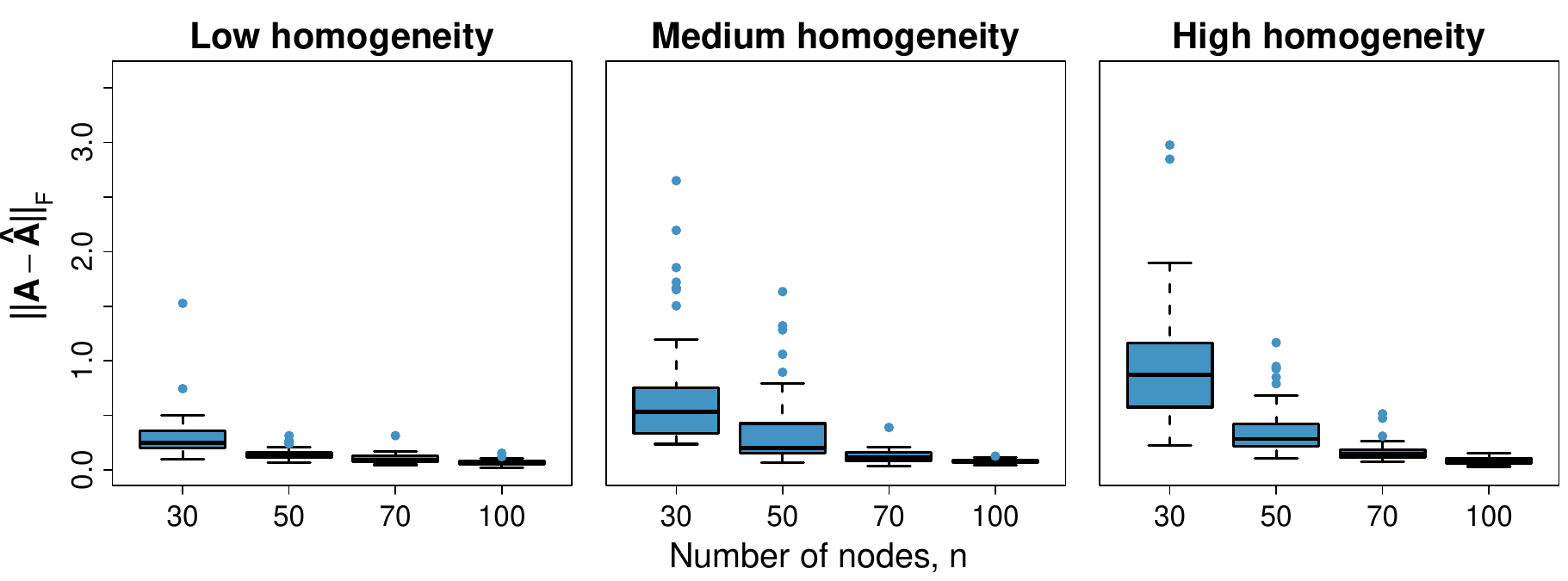}
\subcaption{K=3}\label{param_rec_K=3}
\vspace{1.5mm}
\end{subfigure}
\begin{subfigure}{0.63\textwidth}
\includegraphics[width=1\linewidth,valign=t]{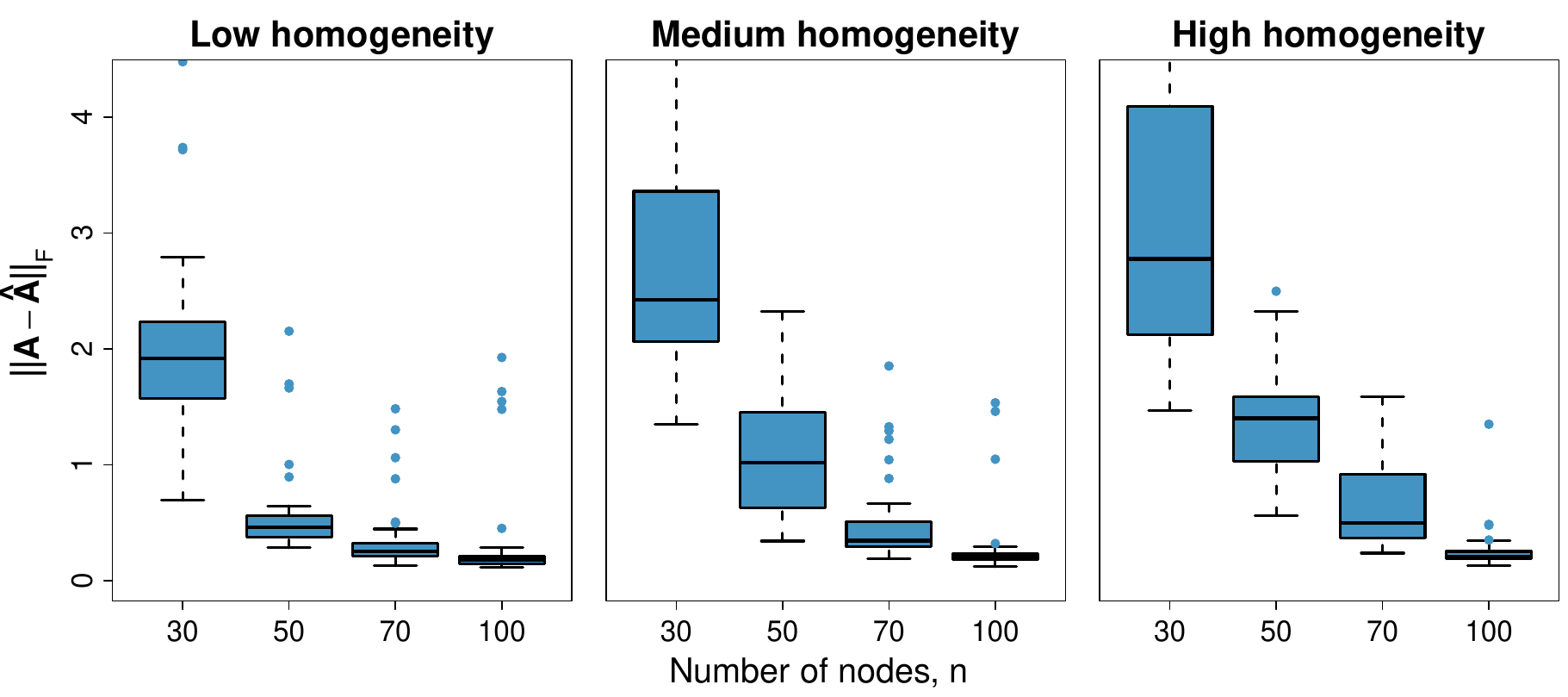}
\subcaption{K=5}\label{param_rec_K=5}
\vspace{-1.5mm}
\end{subfigure}
\caption{Frobenius distance between the true parameter matrix $\mathbf{A}$ and its estimate $\mathbf{\hat{A}}$, based on 50 artificial data sets for varying levels of Dirichlet parameters homogeneity. The number of clusters $K$ is fixed at the respective true values of \subref{param_rec_K=2} 2, \subref{param_rec_K=3} 3 and \subref{param_rec_K=5} 5.}
\label{param_rec_plots}
\end{figure}

The Frobenius distance \citep{matrix_computations} between the true parameter matrix $\mathbf{A}$ and its estimate $\mathbf{\hat{A}}$ is calculated for 50 synthetic data sets in each scenario, and the results are presented in Figure \ref{param_rec_plots}. As expected, larger networks lead to better parameter estimates, showing that the proposed hybrid log-likelihood classification EM leads to consistent parameter estimates as the size of the network increases. Notably, the biggest improvement occurs when we move from $n=30$ to $n=50$. This comes as no surprise since the performance of the model with regards to the recovery of latent clustering improves significantly when the number of nodes increases from 30 to 50 and above, as seen in Section \ref{sim_study_clustering}. We can also note that higher homogeneity of Dirichlet parameters makes the estimation more challenging as a result of higher difficulty in separating the clusters, and the parameter estimates are closer to their true counterparts when the number of clusters is lower. This is again an expected result, as, when the number of clusters increases, recovering the underlying partition of the nodes is more challenging and each individual parameter is estimated from a smaller share of the data.

\subsection{Clustering structure recovery} \label{sim_study_clustering}
\begin{figure}[!tp] 
\centering
\begin{subfigure}{1\textwidth}
\centering
\includegraphics[width=0.7\linewidth,valign=t]{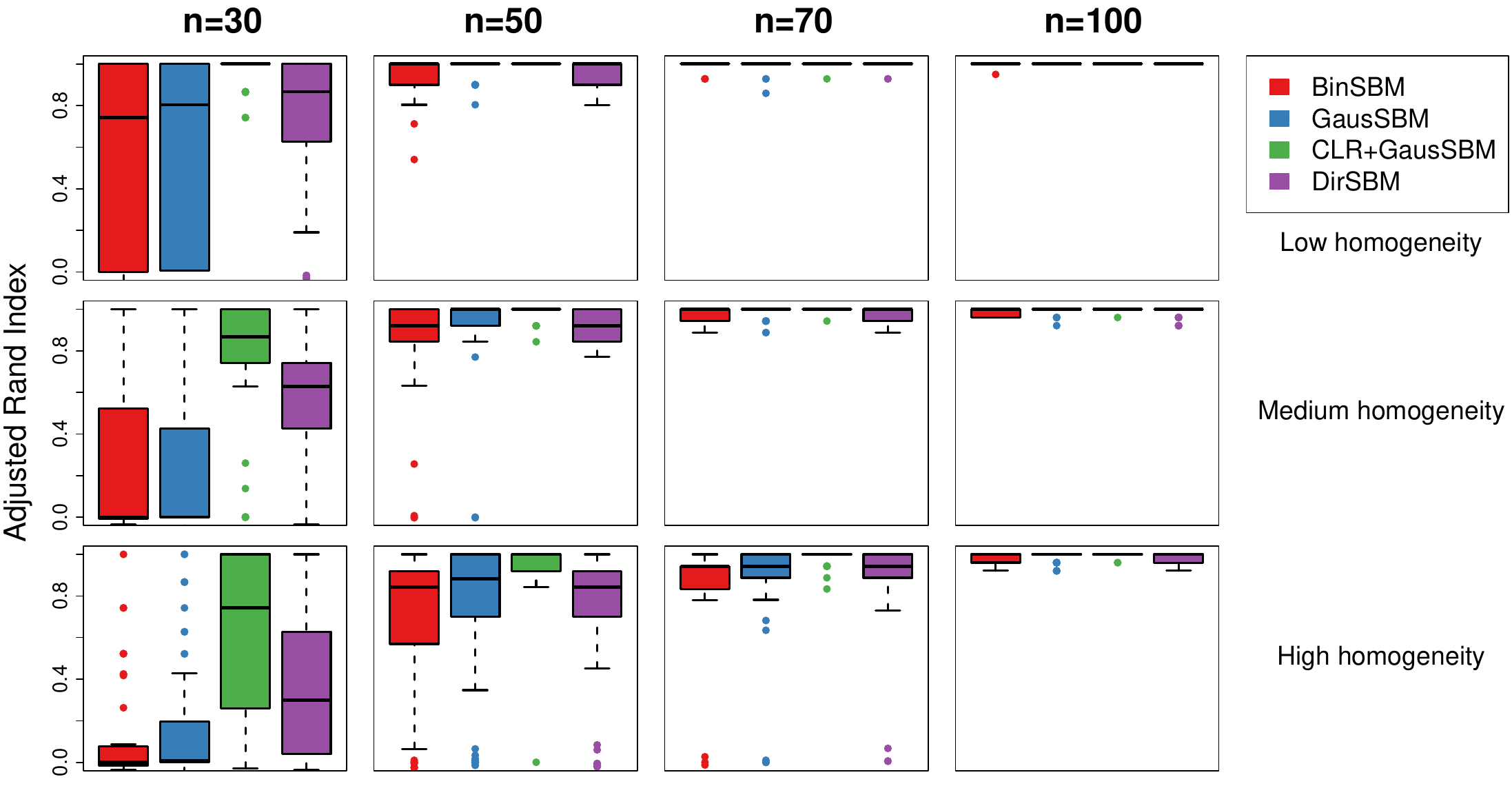}
\subcaption{K=2}\label{sim_study_boxplots_K=2}
\vspace{1mm}
\end{subfigure}
\begin{subfigure}{1\textwidth}
\centering
\includegraphics[width=0.7\linewidth,valign=t]{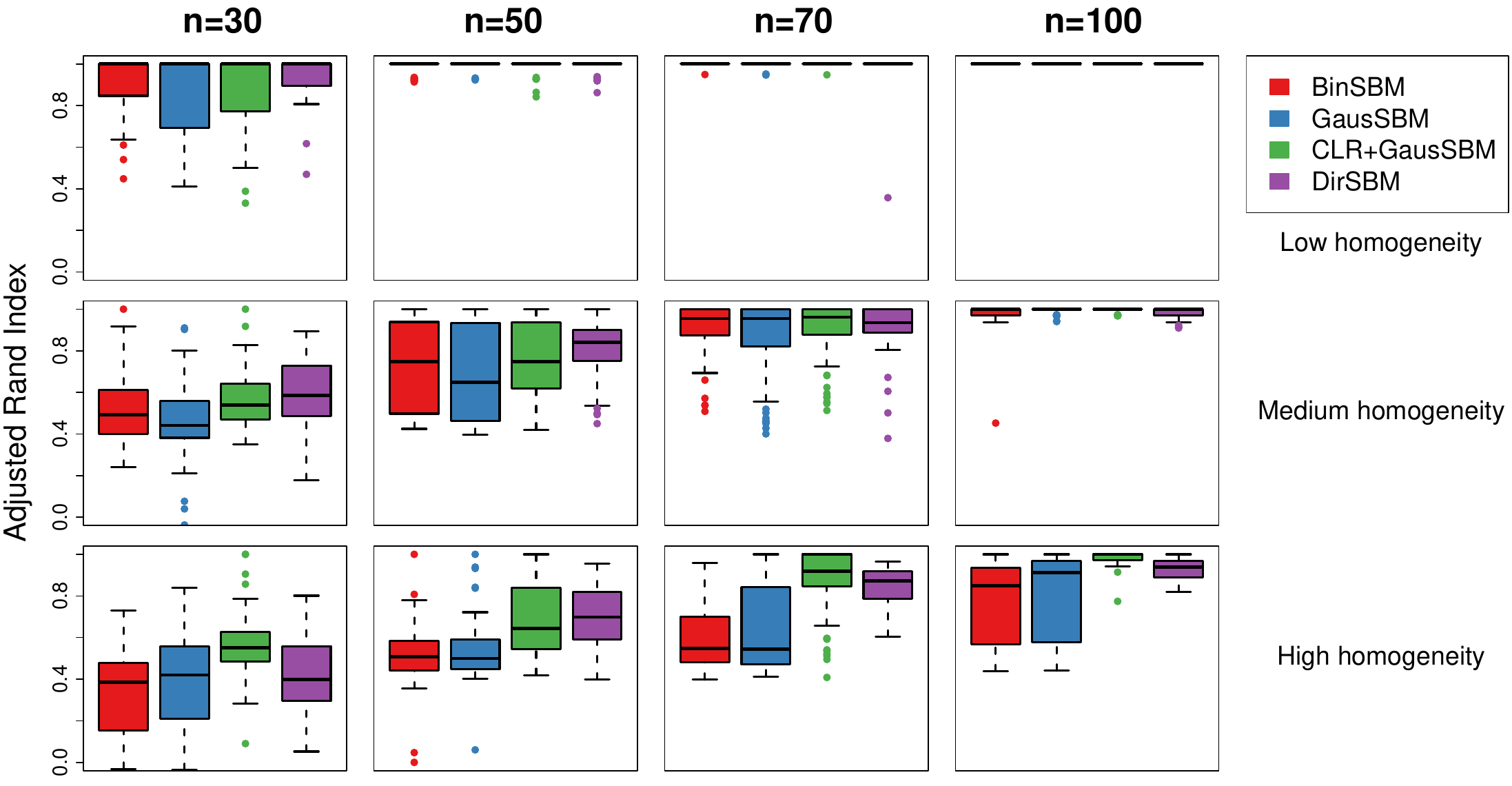}
\subcaption{K=3}\label{sim_study_boxplots_K=3}
\vspace{1mm}
\end{subfigure}
\begin{subfigure}{1\textwidth}
\centering
\includegraphics[width=0.7\linewidth,valign=t]{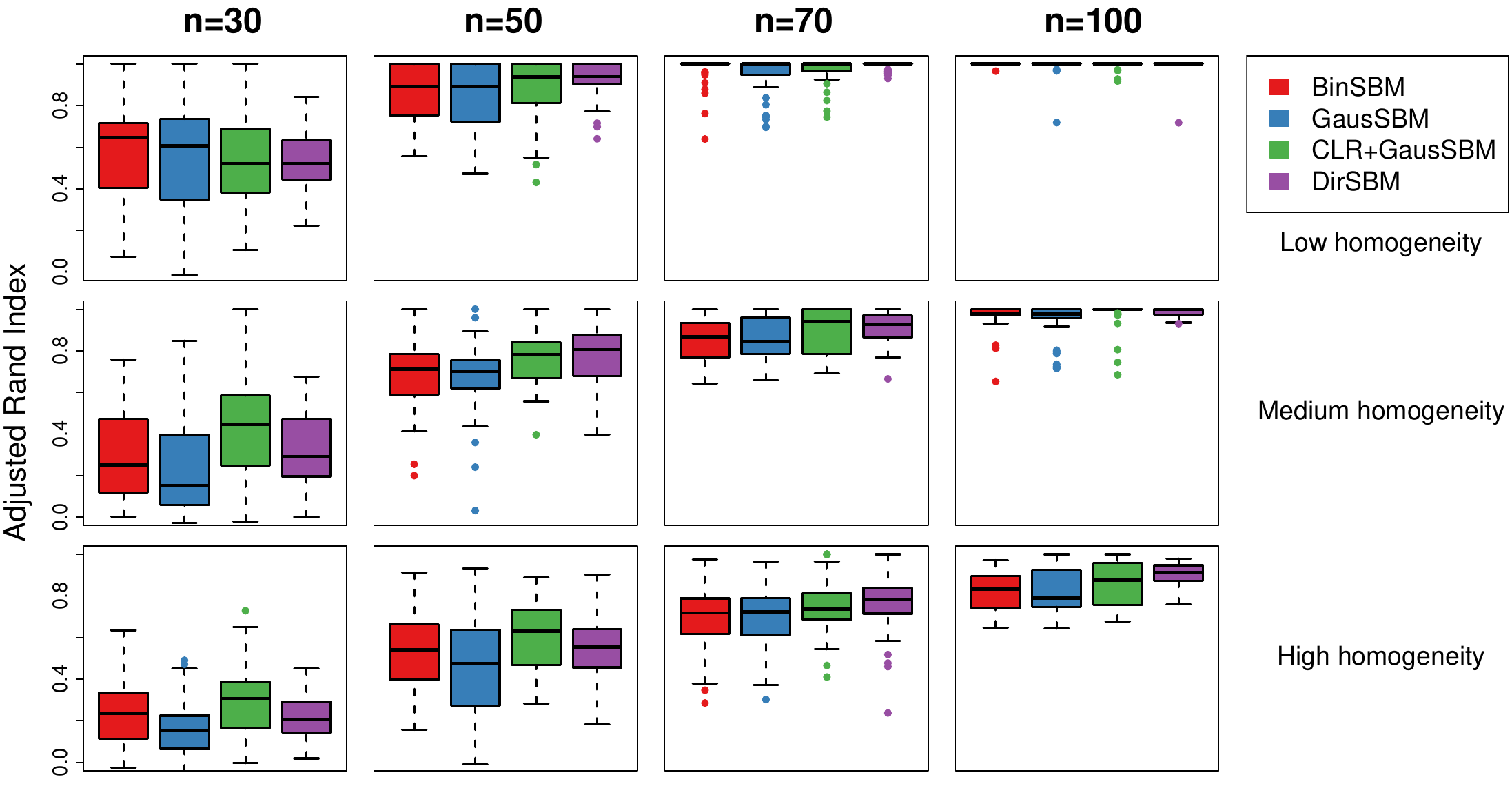}
\subcaption{K=5}\label{sim_study_boxplots_K=5}
\vspace{-2mm}
\end{subfigure}
\caption{Adjusted Rand index (ARI) between the true partition and clustering solutions estimated by binary SBM, Gaussian SBM, Gaussian SBM on centered log-ratio transformed data and DirSBM, with different network sizes $(n)$ and levels of Dirichlet parameters homogeneity. Results are based on 50 artificial data sets with \subref{sim_study_boxplots_K=2} $K=2$, \subref{sim_study_boxplots_K=3} $K=3$ and \subref{sim_study_boxplots_K=5} $K=5$ clusters.}
\label{sim_study_boxplots}
\end{figure}

To assess the quality of the clustering results, we compare the clustering performance of the following models using the ARI:
\begin{enumerate}
    \item \textbf{Binary SBM fitted on the binary version of the network (BinSBM):} \\
    We test the approach that is often taken in practice where a weighted network is transformed into a binary one. Following closely the description of using binary SBM as an initialisation strategy from Section \ref{initialisation}, we construct a binary network and fit the Bernoulli SBM using the $blockmodels$ package in R \citep{blockmodels_package}. 
    \item \textbf{Naive Gaussian SBM (GausSBM):}
    \\
    We fit SBM with Gaussian edge weights \citep{Aicher_weighted_sbm} directly to the data. This model is used to test the robustness to whether accounting for the constant-sum constraint on the sender edge weights or not in different scenarios, as similar naive approaches are sometimes implemented in applications.
    \item \textbf{Gaussian SBM on log-ratio transformed data (CLR+GausSBM):} \label{clr_gaus_sbm}\\
    The data are mapped onto the real line using a centered log-ratio transformation, then SBM with Gaussian weights is fitted to the transformed data, as described in Section \ref{alternative_approach}.
    \item \textbf{DirSBM (DirSBM):}
    \\
    We fit the DirSBM described in Section \ref{Dirichlet_sbm} directly to the data.
    \end{enumerate}

Figure \ref{sim_study_boxplots} illustrates the performance of all 4 models on artificial data sets with 2, 3 and 5 clusters across multiple network sizes and varying levels of Dirichlet parameter homogeneity. Each boxplot is based on 50 networks. The DirSBM was initialised with 5 random cluster partitions based on the results of the simulation study in Section \ref{sim_study_init}. It is evident that the performance of all models improves with sample size, often approaching an ARI of $1$ as we reach 100 nodes. In the case of BinSBM and GausSBM, this can be due to the fact that, in larger networks, the compositional samples are rather high-dimensional, hence the constant-sum constraint does not induce as much dependence between the individual dimensions as in smaller networks, resulting in a good fit of the models that assume independence between the weights. In the case of CLR+GausSBM and DirSBM, the improvement of performance with sample size can be associated with more accurate parameter estimation. Smaller networks present a greater challenge, due to both stronger dependence between the dimensions of compositions and the limited amount of data available for parameter estimation.

The CLR+GausSBM and DirSBM tend to outperform the competitors in terms of clustering structure recovery, and the gap in performance is significantly wider for smaller networks and higher homogeneity of Dirichlet parameters. It is also notable that the performance of BinSBM and the GausSBM on the raw compositional data often exhibits much higher variability indicated by wider boxplots. With regards to the comparison between the DirSBM and the CLR+GausSBM, the two methods tend to perform similarly well. Looking at Figures \ref{sim_study_boxplots_K=3} and \ref{sim_study_boxplots_K=5}, we observe that in many scenarios the DirSBM performs better and with lower overall variability, making it a more a suitable model in the cases with larger number of clusters in the data. Considering Figure \ref{sim_study_boxplots_K=2}, where the number of underlying clusters is 2, we note that CLR+GausSBM tends to perform better than DirSBM. Albeit a simplistic approach, CLR+GausSBM seems to present a valid alternative when modelling compositional networks if it is expected that the number of clusters is small.

\subsection{Model selection performance} \label{sim_study_K}

% \cellcolor{} to color cells
\setlength{\extrarowheight}{2pt}
\setlength{\tabcolsep}{7pt}
\begin{table}[t]
    \caption{Choice of number of clusters using integrated completed likelihood (ICL). Results indicate the number of times each number of clusters has been selected by the criterion, based on 50 artificial networks.}
    \label{tab_model_selection}
    \centering
    \begin{tabular}{cc cccc@{\hskip 20pt} cccc@{\hskip 20pt} cccccc}
    \toprule
        $K$ & & 2 & & & & 3 & & & & 5 & & & & & \\
        \midrule
        $\hat{K}$ & & 1 & 2 & 3 & 4 & 1 & 2 & 3 & 4 & 1 & 2 & 3 & 4 & 5 & 6 \\
        \midrule
        %\hline
        Low homogeneity & $n =$ 30 & 1 & \textbf{45} & 3 & 1 &  & 3 & \textbf{46} & 1 & & 2 & 30 & 18 &  &  \\
        & \textcolor{white}{n = }50 &  & \textbf{50} &  &  &  &  & \textbf{50} &  &  &  &  & 10 & \textbf{40} &  \\ 
        & \textcolor{white}{n = }70 &  & \textbf{48} & 1 & 1 &  &  & \textbf{50} &  &  &  &  & 1 & \textbf{49} &  \\
        & \textcolor{white}{n = }100 &  & \textbf{45} & 5 &  &  &  & \textbf{50} &  &  &  &  &  & \textbf{45} & 5 \\
        \midrule
        Medium homogeneity & $n =$ 30 &  & \textbf{44} & 6 & &  & 37 & \textbf{13} &  & 1 & 20 & 27 & 2 &  &  \\
        & \textcolor{white}{n = }50 &  & \textbf{47} & 3 & & & 10 & \textbf{40} &  &  &  & 3 & 37 & \textbf{10} &  \\ 
        & \textcolor{white}{n = }70 &  & \textbf{47} & 3 & &  &  &  \textbf{48} & 2 &  &  &  & 19 & \textbf{31} &  \\
        & \textcolor{white}{n = }100 &  & \textbf{42} & 7 & 1 &  &  & \textbf{49} & 1 &  &  &  & 2 & \textbf{47} & 1 \\
        \midrule
        High homogeneity & $n =$ 30 & 21 & \textbf{28} & 1 & & 1 & 35 & \textbf{14} &  & 9 & 25 & 16 &  &  &  \\
        & \textcolor{white}{n = }50 & 1 & \textbf{41} & 8 & & & 15 & \textbf{35} &  &  & 1 & 31 & 18 &  &  \\ 
        & \textcolor{white}{n = }70 &  & \textbf{45} & 5 & &  & 2 & \textbf{47} & 1 &  &  &  & 40 & \textbf{10} &  \\
        & \textcolor{white}{n = }100 &  & \textbf{46} & 4 & & &  & \textbf{49} & 1 &  &  &  & 7 & \textbf{43} &  \\
        \bottomrule
         \end{tabular}
\end{table}

We assess the model selection performance using the integrated completed likelihood (ICL) detailed in Section \ref{model_selection}. Table \ref{tab_model_selection} reports the confusion matrices of the true number of clusters $K$ and the optimal number of clusters $\hat{K}$ as selected by ICL. We observe that ICL is successful in selecting the correct number of clusters in larger networks with any level of parameter homogeneity, which is to be expected provided that both the latent structure recovery and the parameter estimation improve with network size. It is also not surprising that, in smaller networks with higher levels of homogeneity and/or higher numbers of clusters, the model selection becomes a much more complex task, and the number of clusters tends to be underestimated by the ICL. This is due to the fact that in cases with medium and high levels of homogeneity the clusters are much harder to separate, coupled with the difficulty of accurately estimating the parameters in smaller networks, as seen in Section \ref{sim_study_clustering} and \ref{sim_study_params} respectively.

\subsection{Effect of zero-weighted edges} \label{zero_sim_study}

In this section we examine the effect of the presence of zero-weighted edges in the network on parameter estimation and clustering performance of DirSBM. To generate synthetic networks with compositional weights arising from weighted networks having a certain proportion $p_0$ of zero-weighted edges and a block structure, we consider the following data generating process: 

\begin{enumerate}
    \item Given a $K$-dimensional vector of cluster membership proportions $\boldsymbol{\theta}$, generate binary cluster allocations $\mathbf{z}_i \sim Multinom(1,\boldsymbol{\theta})$, for $i=1,\ldots,n$. The entries in these vectors are $z_{ik}=1$ when node $i$ is assigned to cluster $C_k$ and 0 otherwise. Denote with $\mathbf{Z}$ an $n\times K$ matrix of binary cluster allocations of all nodes in the network, and with $\mathbf{Z}_{-i}$ the $(n-1)\times K$-dimensional matrix of cluster allocations of all nodes in the network that are not $i$. 
    \item Let $\mathbf{A}=\{\alpha_{kh}\}_{k,h=1}^K$ be a $K \times K$ matrix with strictly positive values. Given the cluster allocations, generate independent gamma realisations corresponding to the raw data
    $$
    y_{ij}\lvert z_{ik}=1, ~ z_{jh}=1 \sim Gamma (\alpha_{kh}, 1)
    $$
    for $i=1,\ldots,n$ and $j=1,\ldots,i-1,i+1,\ldots,n$. 
    \item Set a proportion $p_0$ of edges $y_{ij}$ equal to zero. Consequently, the resulting weighted adjacency matrix $\mathbf{Y}$ will contain a certain percentage of zero-weighted edges.
    \item Set the zero-weighted edges equal to some small quantity $\epsilon$. Here we consider $\epsilon=0.001$. 
    \item From the resulting raw data $\mathbf{Y}$, construct compositions
    $$
        \mathbf{x}_i^* = \bigg( \frac{y_{i1}}{\sum_{j\neq i}^{n}y_{ij}},  \frac{y_{i2}}{\sum_{j\neq i}^{n}y_{ij}},\ldots,\frac{y_{i(i-1)}}{\sum_{j\neq i}^{n}y_{ij}},\frac{y_{i(i+1)}}{\sum_{j\neq i}^{n}y_{ij}},\ldots,\frac{y_{in}}{\sum_{j\neq i}^{n}y_{ij}} \bigg).
    $$
\end{enumerate}

The data generating process above is based on the standard relationship between the Dirichlet distribution and a collection of independent gamma variables: if $y_{ij} \sim Gamma (\alpha_{ij},1)$ for $i=1,\ldots,n$ and $j=1,\ldots,i-1,i+1,\ldots,n$, then the compositions 
$$
\mathbf{x}_i^* = \bigg( \frac{y_{i1}}{\sum_{j\neq i}^{n}y_{ij}},  \frac{y_{i2}}{\sum_{j\neq i}^{n}y_{ij}},\ldots,\frac{y_{i(i-1)}}{\sum_{j\neq i}^{n}y_{ij}},\frac{y_{i(i+1)}}{\sum_{j\neq i}^{n}y_{ij}},\ldots,\frac{y_{in}}{\sum_{j\neq i}^{n}y_{ij}} \bigg) \sim Dir (\alpha_{i1}, \alpha_{i2}, \ldots, \alpha_{i(i-1)},\alpha_{i(i+1)},\ldots,\alpha_{in}).
$$

We consider networks of sizes $n=\{30,50,70,100\}$ with $K=3$ clusters and low and medium level of parameter homogeneity, as defined in Section~\ref{sim_studies} and Appendix~\ref{sim_study_notes}. To better understand how well the model performs when some proportion of the edges is zero-weighted, we test scenarios with $p_0=\{0, 0.01,0.05,0.1,0.25\}$. Similar to the simulation studies in Section \ref{sim_study_params} and \ref{sim_study_clustering}, we use the Frobenius distance between the true Dirichlet concentration parameter matrix $\mathbf{A}$ and its estimate $\hat{\mathbf{A}}$ to assess the parameter estimation performance, and ARI between the true and the estimated partition for clustering performance. 

\begin{figure}[!t] 
\centering
\begin{subfigure}{0.9\textwidth}
\includegraphics[width=1\linewidth,valign=t]{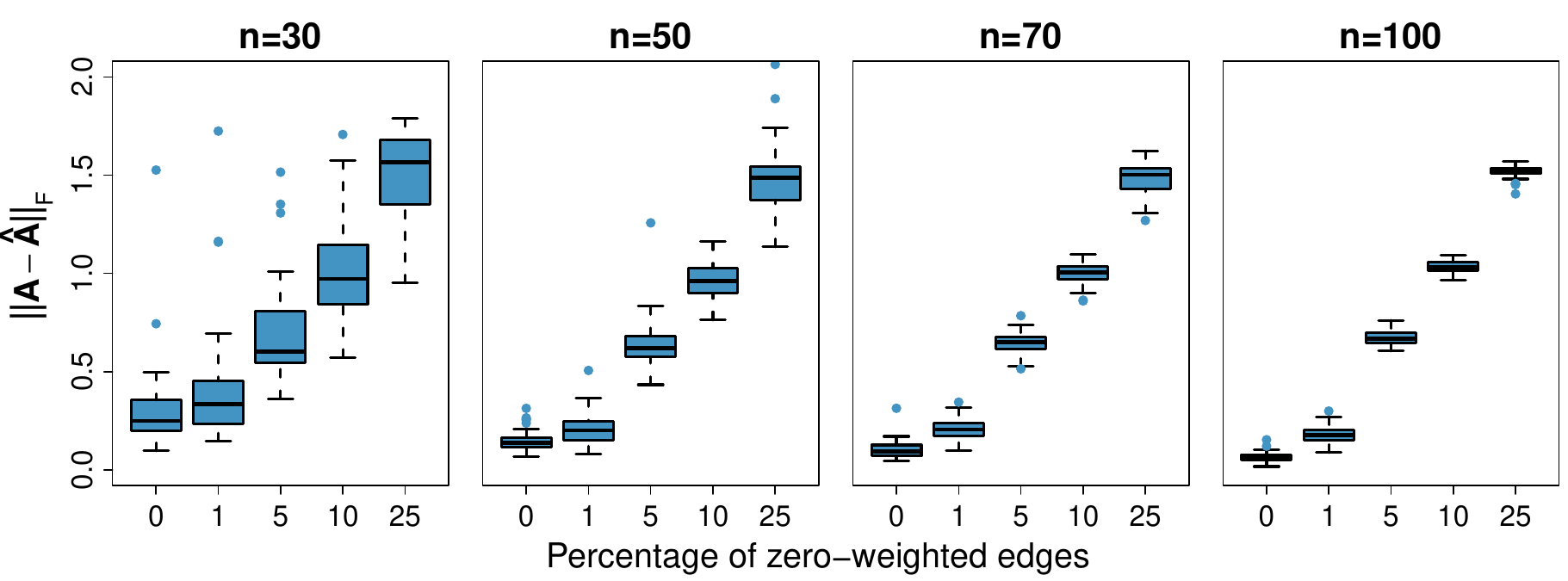}
\subcaption{Low homogeneity} \label{small_zero_weights_params}
\vspace{1.5mm}
\end{subfigure}
\begin{subfigure}{0.9\textwidth}
\includegraphics[width=1\linewidth,valign=t]{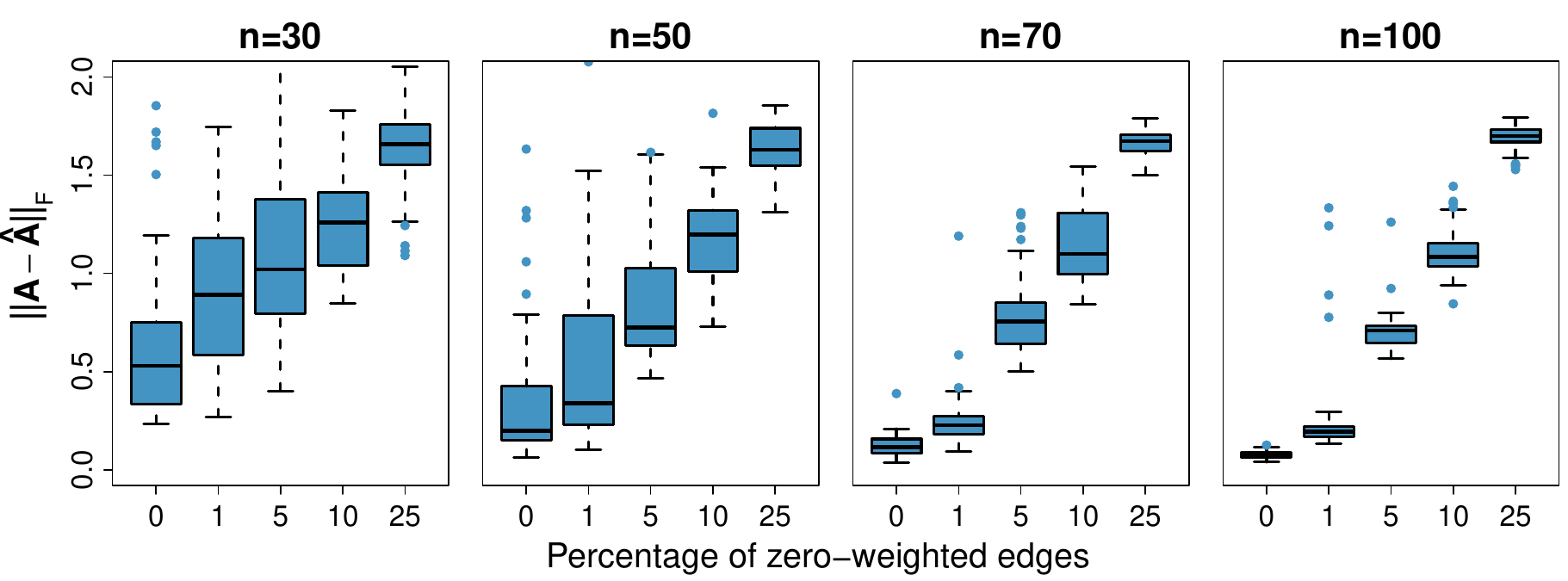}
\subcaption{Medium homogeneity}\label{medium_zero_weights_params}
\vspace{1.5mm}
\end{subfigure}
\caption{Assessment of parameter estimation performance of DirSBM using the Frobenius distance between the true parameter matrix $\mathbf{A}$ and its estimate $\mathbf{\hat{A}}$ for networks with varying percentage of zero-weighted edges, based on 50 artificial data sets with $K=3$ and \subref{small_zero_weights_params} low and \subref{medium_zero_weights_params} medium level of Dirichlet parameter homogeneity.}
\label{zero_weights_params}
\end{figure}

\begin{figure}[!t] 
\centering
\begin{subfigure}{0.9\textwidth}
\includegraphics[width=1\linewidth,valign=t]{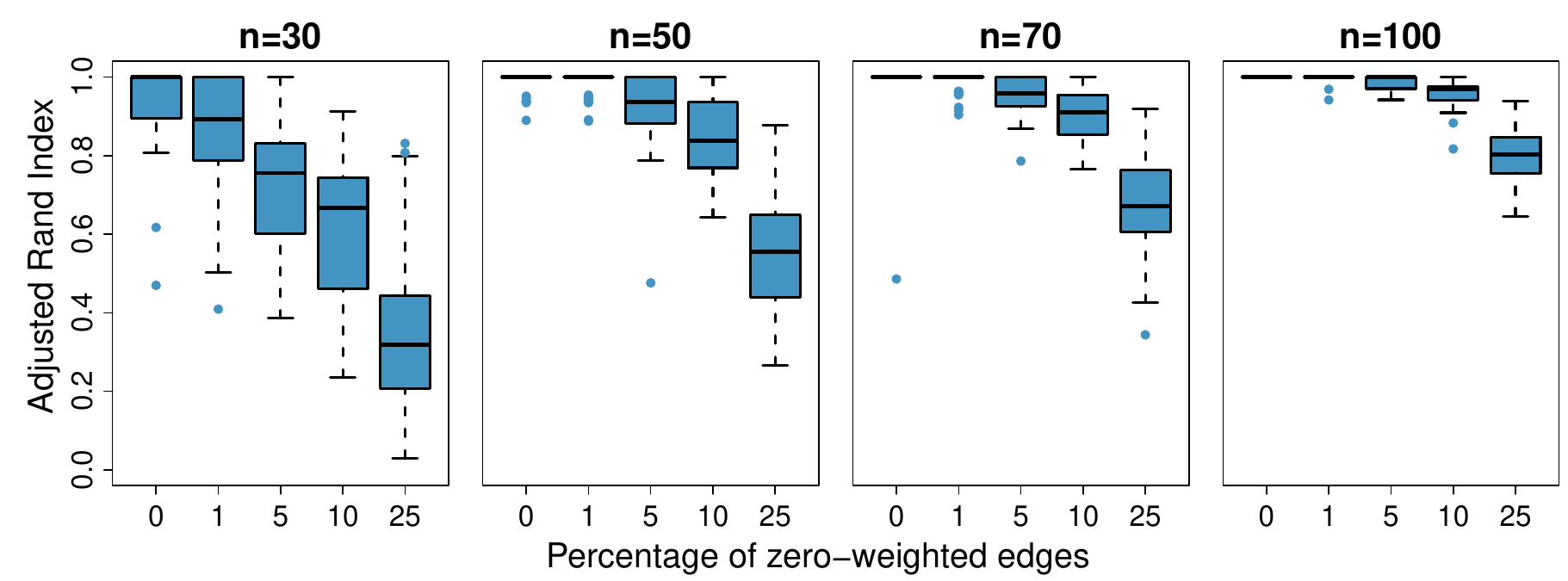}
\subcaption{Low homogeneity} \label{small_zero_weights_ari}
\vspace{1.5mm}
\end{subfigure}
\begin{subfigure}{0.9\textwidth}
\includegraphics[width=1\linewidth,valign=t]{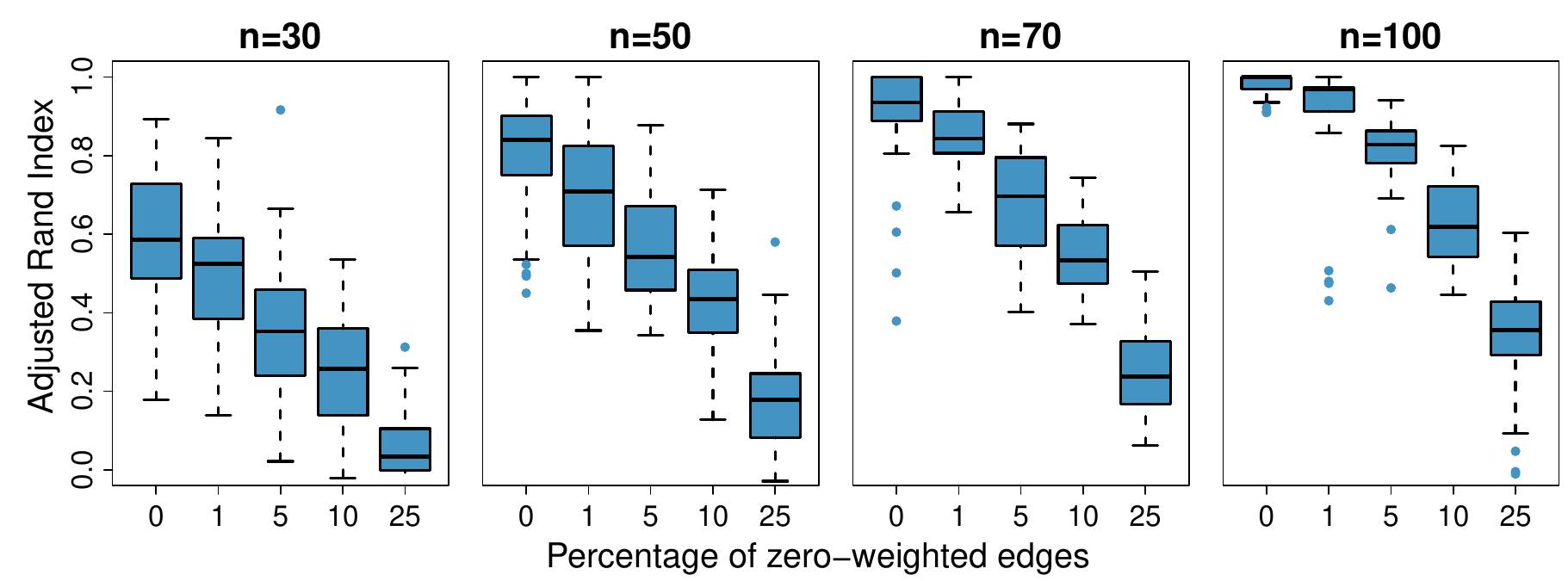}
\subcaption{Medium homogeneity}\label{medium_zero_weights_ari}
\vspace{1.5mm}
\end{subfigure}
\caption{Assessment of clustering performance of DirSBM using the adjusted Rand index (ARI) between the true and the estimated partition for networks with varying percentage of zero-weighted edges, based on 50 artificial data sets with $K=3$ and \subref{small_zero_weights_ari} low and \subref{medium_zero_weights_ari} medium level of Dirichlet parameter homogeneity.}
\label{zero_weights_ari}
\end{figure}

The results are provided in Figure \ref{zero_weights_params} and Figure \ref{zero_weights_ari} respectively, where each boxplot is based on 50 synthetic networks. In Figure \ref{zero_weights_params}, we observe that as the proportion of zero-weighted edges increases, the accuracy of parameter estimates deteriorates on average. This result is expected since, when a higher percentage of zeros is present, fewer informative non-zero weights contribute to the parameter estimation, leading to reduced accuracy. Additionally, we note that the impact of zero-weighted edges on parameter estimation is more moderate in cases with low homogeneity compared to cases with medium homogeneity.

Referring to Figure \ref{zero_weights_ari}, we observe that there is a simultaneous reduction in the quality of the clustering structure recovery as the percentage of zero-weighted edges increases, which is expected as poorer parameter estimation makes it more challenging to correctly partition the nodes. Nonetheless, the reduction is quite gradual, and the performance with a small percentage of zero-weighted edges is close to that with no zero-weighted edges (see also Section~\ref{sim_study_clustering}), in particular in larger networks. The clustering performance of the model appears to be robust to increasing percentage of zero-weighted edges in larger networks with low parameter homogeneity. It is also worth noting that the zero-weighted edges are drawn uniformly across all pairs of cluster labels, meaning that they do not inform the partition as absent edges might in other stochastic block models.

\section{Application to real world network data}
\label{real_data}
\subsection{Erasmus programme data} \label{erasmus_application}
The Erasmus programme is the most popular university student exchange programme in Europe, spanning thousands of Higher Education institutions in more than 30 countries. \cite{Erasmus_data} have produced an extensive data set of student, staff and teacher mobility in the Erasmus network by combining data from a number of different sources. The aim of our analysis is to understand global mobility patterns in the Erasmus network, focusing on student exchanges. Due to significant differences in the student populations of the European countries, analysis of raw counts of students per pair of countries may not provide particular insights regarding the popularity of destinations, as the biggest countries dominate simply due to their overall hosting capacity. 

After aggregating the student numbers to country level for the 2012-2013 academic year, a directed network with countries as nodes and proportions of students as edge weights is constructed, with the proportions summing to 1 for the sending countries. The Erasmus network is very dense ($\approx 87\%$ of all possible edges are present), and there is no reason to believe that the remaining edges are not allowed to exist since all countries in the Erasmus network are allowed to exchange students. Therefore, instead of treating the edges as absent, and as discussed in Section~\ref{zero-weights}, we assign a value of $0.001$ to the interactions with zero count and treat the network as fully connected. 

We implement DirSBM with $K$ ranging from 1 to 6 and use the ICL to select the number of clusters. 
According to the ICL, the optimal number of clusters is 3 (ICL$=3764$), followed by a solution with 5 clusters (ICL$=3718$). As the Erasmus network is fairly small (33 nodes), the ICL might underestimate the number of clusters due to the penalty being too strong in the context of small $n$, as we saw in Section \ref{sim_study_K}, therefore we examine the results of both $K=3$ and $K=5$ cases.

\begin{figure}[tp] 
\centering
\begin{subfigure}{1\textwidth}
\includegraphics[width=1\linewidth,valign=t]{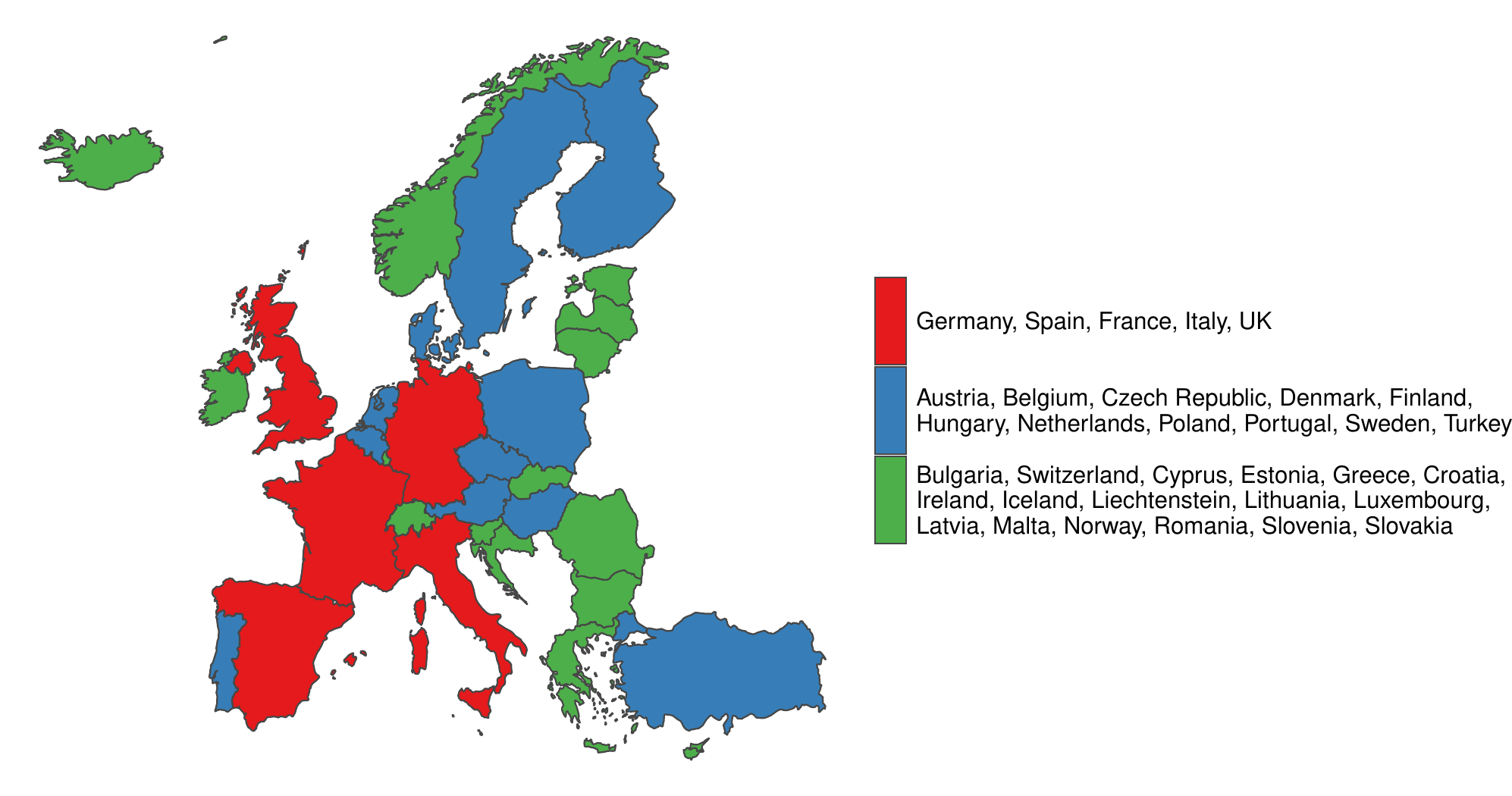}
\subcaption{}
\label{Maps_sols_K=3}
\end{subfigure}
\begin{subfigure}{1\textwidth}
\centering
\includegraphics[width=0.6\linewidth,valign=t]{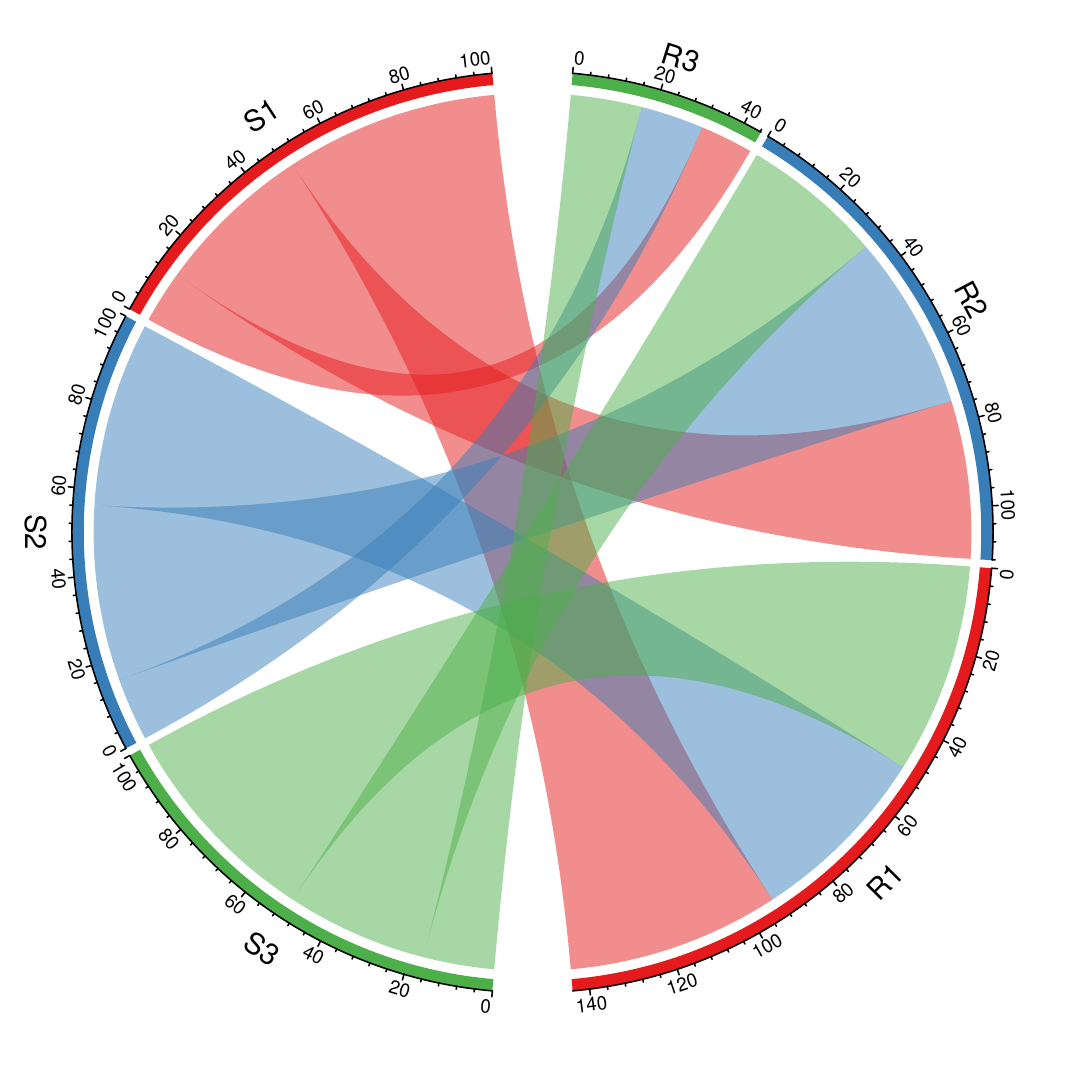}
\subcaption{}
\label{Erasmus_sols_K=3}
\end{subfigure}
\caption{Clustering solution for DirSBM with 3 clusters: \subref{Maps_sols_K=3} map of Europe with countries coloured by cluster assignment; \subref{Erasmus_sols_K=3} chord diagram of total percentage flows between clusters of countries based on $\mathbf{\hat{V}}$. On the left, S1, S2 and S3 denote the sending cluster (1, 2 and 3). Percentages departing from the sender sum up to 100. On the right, R1, R2 and R3 denote the receiving cluster (1, 2 and 3).}
\label{Erasmus_K=3}
\end{figure}

Figure \ref{Maps_sols_K=3} illustrates the solution with 3 clusters on a map of Europe, noting the membership of each cluster on the side as smaller countries are difficult to see on the plot. Figure \ref{Erasmus_sols_K=3} represents the total percentage flows between the clusters, based on the estimated matrix $\mathbf{\hat{V}}$. The estimate of the parameter matrix $\mathbf{A}$ as well as the expected node-to-node and cluster-to-cluster exchange matrices $\mathbf{\hat{W}}$ and $\mathbf{\hat{V}}$ from Equations \eqref{W_matrix} and \eqref{V_matrix} respectively are provided in \ref{estimates_erasmus}. 

The clustered compositions are reasonably interpretable in a real-world context. The red group seems to dominate the student preference list across all clusters as indicated by high country-wise exchange percentages (12.6, 8.7 and 10 respectively) from the first column of $\mathbf{\hat{W}}$, and it consists of Germany, Spain, France, Italy and the UK. The red cluster collectively also notably receives the highest total percentages. The blue cluster can be described as containing countries that are still very popular exchange destinations, but not as popular as the likes of Germany and France, possibly due to their relatively small size or more challenging national language. Although roughly equal percentages of students are expected to be retained by the blue cluster and to be sent to the countries in the red cluster, 41.4\% and 43.7\% respectively, these shares are split between 11 countries instead of 5, hence each individual country in the blue cluster is expected to receive smaller shares in comparison to countries in the red cluster; for instance, any red cluster country is expected to receive 10\% of students from any green cluster country, whereas for blue cluster countries this share is only 3\%. The green cluster contains the less popular exchange destinations, such as Bulgaria and Slovakia, as indicated by significantly lower expected country-wise and cluster-wise exchange shares. Collectively, the green cluster also retains the least of students, as only 16.8\% go on exchange within the cluster itself, while the remaining 83.2\% go on exchange to the countries in the red and blue clusters. It can also be noted that the countries in the red cluster are generally closely co-located in geographical terms, as are most countries in the blue cluster, except for Portugal and Turkey, and the countries in the green cluster are more spread out.

\begin{figure}[tp] 
\centering
\begin{subfigure}{1\textwidth}
\includegraphics[width=1\linewidth,valign=t]{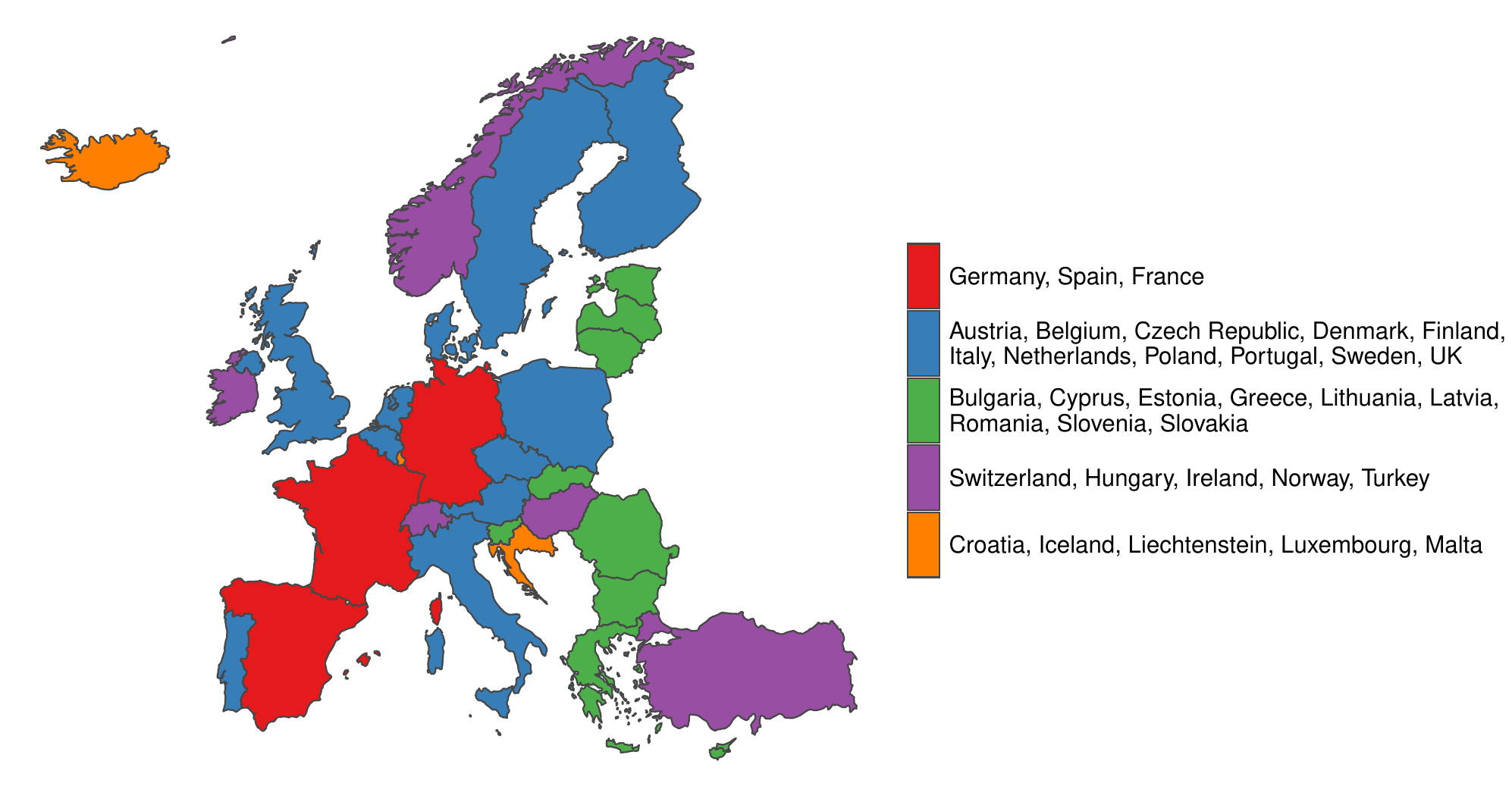}
\subcaption{}
\label{Maps_sols_K=5}
\end{subfigure}
\begin{subfigure}{1\textwidth}
\centering
\includegraphics[width=0.6\linewidth,valign=t]{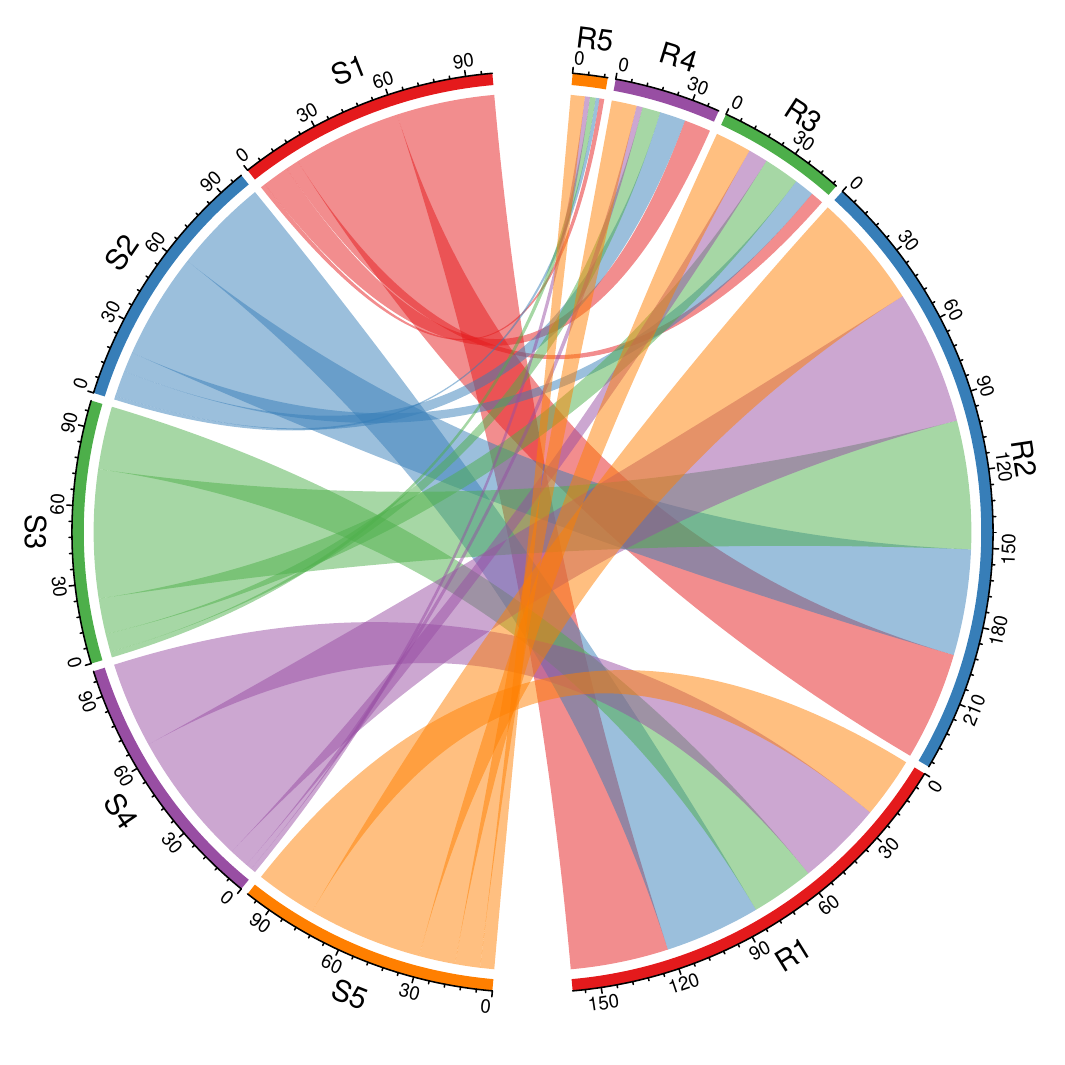}
\subcaption{}
\label{Erasmus_sols_K=5}
\end{subfigure}
\caption{Clustering solution for DirSBM with 5 clusters: \subref{Maps_sols_K=5} map of Europe with countries coloured by cluster assignment; \subref{Erasmus_sols_K=5} chord diagram of total percentage flows between clusters of countries based on $\mathbf{\hat{V}}$. On the left, S1-S5 denote the sending cluster (numbered 1 to 5). Percentages departing from the sender sum up to 100. On the right, R1-R5 denote the receiving cluster (numbered 1 to 5).}
\label{Erasmus_K=5}
\end{figure}

The second best solution according to ICL with $K=5$ is presented in Figure \ref{Erasmus_K=5}, and the parameter estimates as well as matrices $\mathbf{\hat{V}}$ and $\mathbf{\hat{W}}$ are available in \ref{estimates_erasmus}.
The red cluster, now consisting of only Germany, Spain and France, is the cluster of countries receiving the highest percentages of students from the countries of all clusters, ranging between the expected 8.1\% to 19.6\% per country pair, and the second largest total shares, dominated by the blue cluster by a small margin. It should be noted that the total shares received by the red cluster are only split between 3 countries instead of 11 like in the case of the blue cluster. This cluster is also characterised by retaining 39.1\% of its students, and sending 43.3\% to the blue cluster, implying a lack of diversity of exchange destinations among its students. The blue cluster can be described as containing reasonably popular exchange destinations, but not quite as popular as the red countries, and, similarly to the red cluster, the majority of students go on exchange within the cluster itself (42.2\%) or to the red cluster (38.1\%). The green cluster, despite consisting of 9 countries (8 Eastern European countries and Cyprus), does not attract large shares of students (total cluster exchange shares are between 2.4\% and 10.6\%), which may be due to the fact that such countries are less well known exchange destinations. Green countries have a somewhat more diverse preference profile in comparison to the first two clusters, as the parameter estimates are much closer to each other, but there is still a strong preference for the red and blue cluster countries. Purple countries, which are Switzerland, Hungary, Ireland, Norway and Turkey, can be described as medium-preference exchange destinations. They have a strong preference for the red and blue cluster countries, and very limited interest in countries within its own cluster, as well as in green and orange clusters. The orange cluster is the one having the most homogeneous preferences, as indicated by the smallest differences between the parameter estimates, and its countries tend to be among the least popular destinations for exchange. This can possibly be explained by the fact that these countries are either very small, are located very far from the centre of Europe or are not members of the European Union (as of the 2012-2013 academic year). The green, purple and orange clusters retain very small shares or transfers, 14.2\%, 2.4\% and 5.6\% respectively, while the majority of students from these clusters tend to go on exchange to countries in the red and blue clusters. A final observation is that there seems to be a connection between the exchange preferences of Erasmus students and the level of economic development of the hosting country or the prestige of going on exchange to specific destinations. The countries with stronger economies tend to be grouped together and to exhibit higher expected country-to-country exchange proportions, whereas countries with weaker economies or newer members of the programme, such as Turkey and Croatia, tend to be separated from the more established members.

\subsection{London bike sharing data} \label{bike_sharing_application}

We illustrate DirSBM in application to the London bike sharing data from 2014 provided by the \cite{TFL}. We start by counting the number of trips taken between pairs of stations and identifying the busiest stations for our analysis. To do so, we look at the top 100 start and end stations in terms of volume of bikes exchanged and consider their intersection, i.e. the stations that are both among the most popular start and end points of the journey, resulting in an overall network of 85 stations. The trips only between these 85 stations account for over 10\% of the total year's trips. The network is very dense, with just under 2\% of edges having zero weights, hence, similarly to the Erasmus programme network from Section \ref{erasmus_application} and as discussed in Section~\ref{zero-weights}, we set the value of the zero-weighted edges equal to $0.001$. We then compute the compositional edge weights by dividing the count weights by the total number of trips taken from the start station. 

\begin{figure}[tp] 
\centering
\begin{subfigure}{1\textwidth}\
\centering
\includegraphics[width=0.95\linewidth,valign=t]{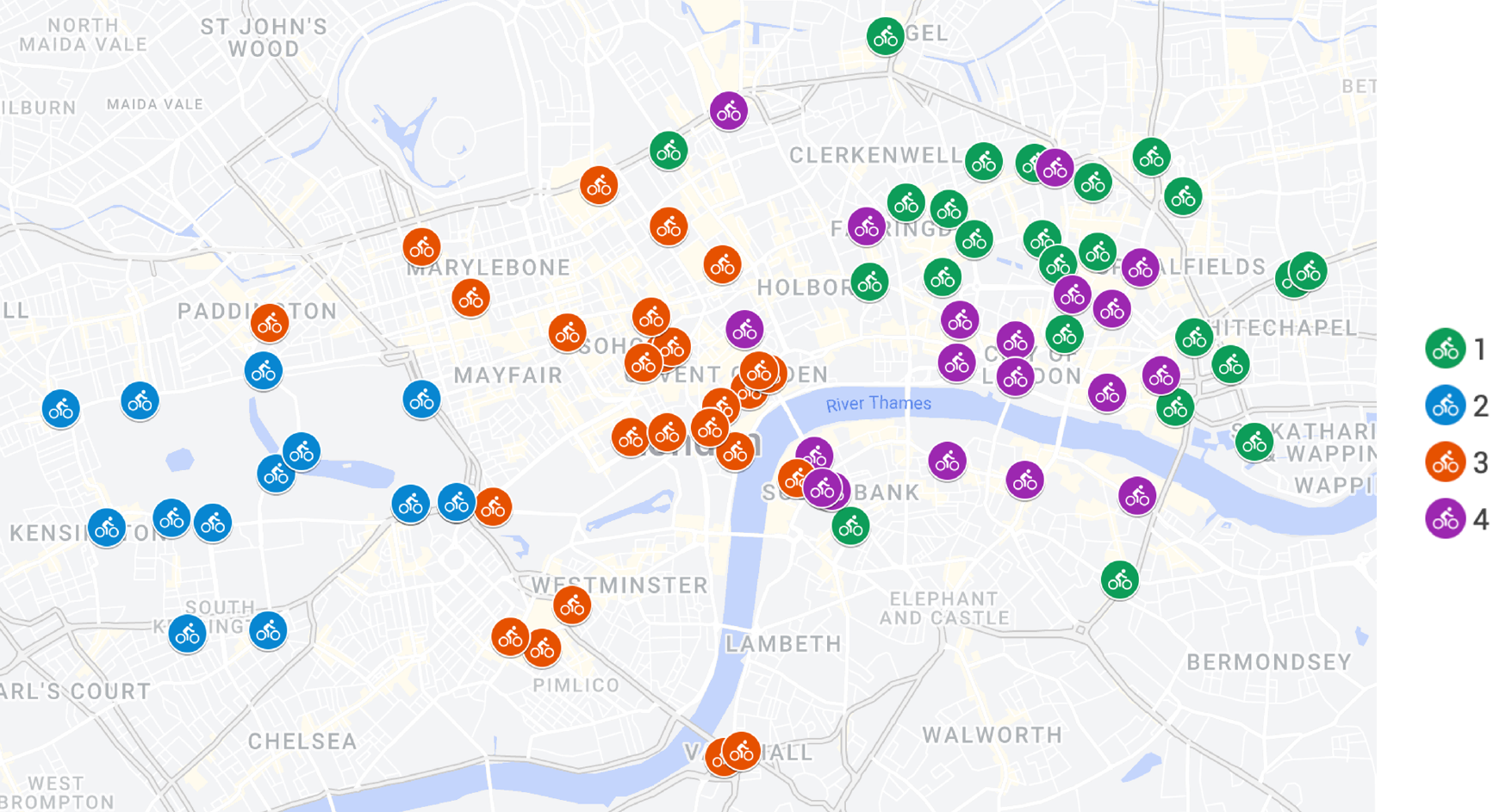}
\subcaption{}
\label{bike_map}
\end{subfigure}
\begin{subfigure}{1\textwidth}
\centering
\includegraphics[width=0.6\linewidth,valign=t]{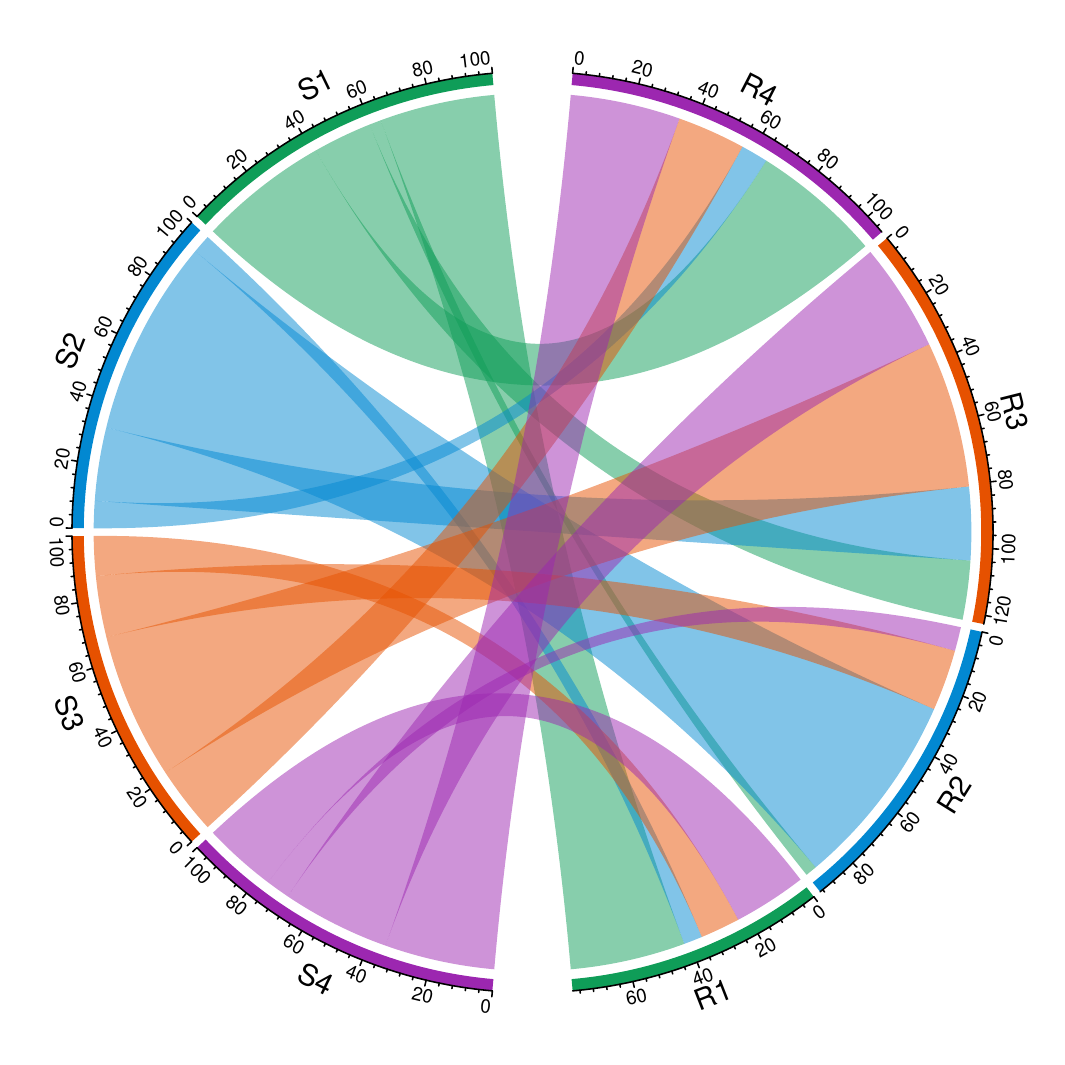}
\subcaption{}
\label{bike_sols_K=4}
\vspace{-2mm}
\end{subfigure}
\caption{Clustering solution for DirSBM with 4 clusters, as selected by ICL, on London bike sharing data:  \subref{bike_map} map of Central London with bike stations coloured by cluster assignment. Image created using \href{https://www.google.com/maps/d/}{Google My Maps} software, with the interactive map available \href{https://www.google.com/maps/d/edit?mid=1Jjv8zwCHdPPvxdK6gBFqmf3vLng04mg&usp=sharing}{here}; \subref{bike_sols_K=4} chord diagram of total percentage flows between clusters of stations based on $\mathbf{\hat{V}}$. On the left, S1-S4 denote the sending cluster (numbered 1 to 4). Percentages departing from the sender sum up to 100. On the right, R1-R4 denote the receiving cluster (numbered 1 to 4).}
\label{bike_sol}
\end{figure}

We fit the DirSBMs with number of clusters between 1 and 8 clusters and select the optimal number of clusters based on the ICL. According to the criterion, the solution with $K=4$ is optimal (ICL$=27866$), followed by the solution with $K=3$ (ICL$=27608$). Figure \ref{bike_sol} illustrates the solution with 4 clusters on a map of central London (an interactive version of the map is available \href{https://www.google.com/maps/d/edit?mid=1Jjv8zwCHdPPvxdK6gBFqmf3vLng04mg&usp=sharing}{here}) as well as the chord diagram for total exchange proportions between clusters. The Dirichlet parameter matrix estimate $\mathbf{\hat{A}}$ as well as matrices $\mathbf{\hat{V}}$ and $\mathbf{\hat{W}}$ are provided in \ref{estimates_bike}.

Unlike the Erasmus network data presented in Section \ref{erasmus_application}, the clustering structure is largely driven by stronger connections within clusters than those between clusters as the diagonal elements of $\mathbf{\hat{A}}$ tend to be dominant. The clusters are also much more balanced in terms of number of observations, with 24, 15, 26 and 20 stations in the green, blue, red and purple cluster, respectively. There is a geographical component to the clustering structure recovered, in the sense that the clusters are quite compact in space, indicating that more trips take place within the neighbouring areas of the city, and the neighboring clusters tend to exchange larger shares of trips in comparison to more distant ones.

The blue cluster is centered around Hyde Park and Kensington, which has the least geographical spread, and it retains the majority of its flow within the cluster (62.3\% of trips), possibly due to the purpose of the journey being recreational cycling. A portion of 23.2\% of trips departing from the blue cluster terminate in the neighboring red cluster, and significantly smaller shares arrive to the purple (8.6\%) or the green clusters (5.9\%).
The red cluster contains bike stations around Soho and Covent Garden regions of London that are known for entertainment and shopping, as well as stations near major tourist attractions (such as the British Museum) or large train stations (e.g. Euston station and King's Cross station). A significant 46.4\% of trips take place within the cluster, and the green cluster receives the least trips starting from the red cluster, possibly due to the geographical distance between the two. The green and the purple clusters have strong connections with each other, exchanging 41.1\% and 23.2\% of their trips (in comparison to retaining 36.5\% and 34.9\% respectively), which can be due to the concentration of corporate buildings in this region of London and the employees living within cycling distance of their workplaces. Neither cluster is well connected to the blue cluster, likely due to the distance and availability of faster public transport options as well as the purpose of travel. A proportion of 34.3\% of trips from stations in a purple cluster arrive in the red cluster, whilst 21.6\% travel in the opposite direction. In summary, the clusters detected by DirSBM by modelling the proportions of transfers between the stations are interpretable in terms of geographical locations and neighbourhoods of London.

\section{Discussion} \label{discussion}

In this paper, we introduced DirSBM, an extension of the stochastic block model to directed weighted networks with compositional edge weights that is based on direct modelling of compositional weight vectors using the Dirichlet distribution. To the best of our knowledge, no counterparts currently exist in the literature, making the proposed clustering methodology a novel contribution. We have developed an inferential procedure based on a variant of the classification expectation-maximisation algorithm for hybrid likelihood. Model selection is addressed using the integrated completed likelihood criterion. We have also proposed an alternative framework for clustering for composition-weighted networks via a log-ratio transformation of the data and subsequent application of an existing weighted SBM to the transformed data. Through simulation studies, we have shown the effectiveness of DirSBM in terms of parameter estimation and recovery of the clustering structure in a variety of settings, and have explored the use of ICL for model selection, which has proven to work reasonably well. The code implementing the modelling framework in R can be found on \href{https://github.com/iuliiapromskaia/dirSBM}{GitHub}.

The modelling approach can be extended in multiple directions. As mentioned at the start of Section \ref{Dirichlet_sbm}, the proposed DirSBM models the network's latent structure based on the sending proportions of the nodes, meaning that the resulting groups are associated with the structure of the nodes acting as senders. However, the latent structure based on the receiving proportions could be quite different as often the nodes behave differently depending on the role they play in the interaction, as noted in \cite{mixed_membership_sbm}. 
Exploring the clustering based on receiving proportions or combining sending and receiving information by extending DirSBM to a two-layer network, with one layer for sending and another for receiving proportions, would be an interesting avenue for future research.

Another important extension to the DirSBM is introducing structural zeros, to enable the ability to model the absence of guaranteed edges in the network. One could use the idea from \cite{dir_reg_with_zeros} that concerns Dirichlet regression with zeros, where indicators are used to ``discard" zero entries of compositional vectors and subsequently fit the model to the non-zero subset of compositional data. As well as making the model more realistic and applicable to a wider range of real world data, such an extension could lead to numerically more stable performance, since the dimension of the Dirichlet observations would not scale with the number of nodes. 

The use of the centered log-ratio transformation approach presented in Section \ref{alternative_approach} has been shown to be a simple valid alternative to the DirSBM in simple scenarios with few clusters in small networks. With regards to this approach, transformations other than the centered log-ratio could be explored as well as other weighted SBMs. As for the incorporation of structural zeros, it is unclear how one could include such information in this methodology as the edges are treated as independent and the respective parameters of the distribution are estimated in the Euclidean space \citep[Chapter 13.5]{coda_book}. This implies that even if two nodes have similar weights in a compositional sense, the absence of edges for one of them can lead to assignment to different clusters, because the absent edges would make them appear more different than they actually are when mapped to the Euclidean space. To give a simple illustrative example, suppose we have two compositions, $x_1=(0.2,0.5,0.28,0.01,0.01)$ and $x_2=(0.2,0.5,0.3,0,0)$, and let the zero values be structural zeros, i.e. indicating a missing edge in the network. Then, applying the centered log-ratio transformation, $u_1=clr(x_1)=(0.95,1.86,1.28,-2.05,-2.05)$ and $u_2=clr(x_2)=(-0.44,0.48,-0.035,0,0)$ (using the convention $\log 0=0$), a false impression is created that the compositions are quite different. 

Other developments could include investigating more educated and efficient initialisation strategies for the proposed algorithm. Although random initialisation allows to explore the solution space rather well by considering the most diverse set of starting points, thus reducing the risk of convergence to a local maximum, it does so at a computational cost. This cost could potentially be reduced if the algorithm was to be run from fewer, or ideally a single, more informed initial clustering allocation set. Additionally, alternative estimation methods could also be explored in an attempt to reduce the run times. For instance, the profile pseudo-likelihood approach as proposed by \cite{pseudo-likelihood} might offer computational advantages. Unlike \cite{HybridML}, where a classification step is implemented within each iteration of the algorithm, in \cite{pseudo-likelihood} the columns of the data matrix are classified only after the EM has converged. While such an approach could enhance speed for most SBMs with conditionally independent edges or edge weights, its efficacy for DirSBM is less certain due to the dependence of individual edge weight vectors on the global network latent structure. Any implementations aiming to significantly reduce the computational burden would need to take into account this characteristic of DirSBM and possibly target the estimation of the Dirichlet concentration parameters. Nonetheless, further investigations in this direction could prove fruitful.

Lastly, other distributions for compositional data could be considered to model the sets of proportional edge weights in the network, such as the generalised Dirichlet distribution, which could be a less restrictive approach to capturing compositional variance \citep{generalised_dirichlet}.

\section*{Acknowledgements} 

This work was supported by the Insight Research Ireland Centre for Data Analytics (Grant number 12/RC/2289\_P2).


\renewcommand{\appendixname}{Appendix}

\appendix
\section{Complete data log-likelihood} \label{l_c}

Recall that given the binary matrix of cluster allocations $\mathbf{Z}$, the compositional data $\mathbf{X}$ has the following probability density:
\begin{equation}
    p(\mathbf{X}| \mathbf{Z}) = \prod_{i=1}^{n} p(\mathbf{x}_i| \mathbf{z}_i,\mathbf{Z}_{-i}) =  \prod_{i=1}^{n} \prod_{k=1}^{K} \Bigg [\frac{\Gamma (\sum_{j\neq i}^{n} \alpha_j)}{\prod_{j\neq i}^{n} \Gamma (\alpha_j)} \prod_{j\neq i}^{n} x_{ij}^{\alpha_j-1} \Bigg ] ^{z_{ik}}, \text{where } \alpha_j=\sum_{h=1}^{K} z_{jh} \alpha_{kh}.
\end{equation}
The latent variables have probability distribution:

\begin{equation}
    p(\mathbf{z}_i)=\prod_{k=1}^{K} \theta_k^{z_{ik}}.
\end{equation}
This results in the following complete data likelihood:

\begin{equation} \label{p(X)}
\mathcal{L}_c(\mathbf{A},\boldsymbol{\theta}) =  \prod_{i=1}^{n} \prod_{k=1}^{K}
p(\mathbf{x}_i,\mathbf{z}_i,\mathbf{Z}_{-i}) =\prod_{i=1}^{n} \prod_{k=1}^{K} \Bigg [ \theta_k \frac{\Gamma (\sum_{j\neq i}^{n} \alpha_j)}{\prod_{j\neq i}^{n} \Gamma (\alpha_j)} \prod_{j\neq i}^{n} x_{ij}^{\alpha_j-1} \Bigg ] ^{z_{ik}}.
\end{equation}
Taking the natural logarithm of Equation \eqref{p(X)}, we arrive at the expression for the complete data log-likelihood from Equation \eqref{Complete_ll}.

\section{Classification EM with hybrid log-likelihood} \label{CEM_appendix}

The expectation of the complete data hybrid log-likelihood with respect to a single latent variable $\mathbf{z}_i$ is given by:

\begin{equation}
\begin{aligned}
    & \mathbb{E} [l_c^{hyb}(\mathbf{A},\boldsymbol{\theta})] \\ & = \sum_{i=1}^{n} \sum_{k=1}^{K} \mathbb{E}[z_{ik}] \Bigg (\log \Gamma (\sum_{j\neq i}^{n} \Tilde{\alpha}_j) - \sum_{j\neq i}^{n} \log \Gamma (\Tilde{\alpha}_j) + \sum_{j\neq i}^{n} (\Tilde{\alpha}_j - 1) \log x_{ij} \Bigg ) + \sum_{i=1}^{n} \sum_{k=1}^{K} \mathbb{E}[z_{ik}] \log \theta_k,    
\end{aligned}
\end{equation}
with $\Tilde{\alpha}_j =\sum_{h=1}^{K} \Tilde{z}_{jh} \alpha_{kh}$, where $\Tilde{z}_{jh}$ is the fixed cluster indicator $\Tilde{z}_{jk} = \mathbbm{1}\{\Tilde{c}_j=k\}$.

The E-step can be found in standard EM fashion by using the Bayes rule:

\begin{equation}
\begin{aligned}
    \hat{z}_{ik} = \widehat{\Pr}(z_{ik}=1|\mathbf{x}_i,\widetilde{\mathbf{Z}}_{-i})&=\dfrac{p(z_{ik}=1|\widetilde{\mathbf{Z}}_{-i})p(\mathbf{x}_i|z_{ik}=1,\widetilde{\mathbf{Z}}_{-i})}{p(\mathbf{x}_i|\widetilde{\mathbf{Z}}_{-i})} \\[1ex] &  =\dfrac{p(z_{ik}=1)p(\mathbf{x}_i|z_{ik}=1,\widetilde{\mathbf{Z}}_{-i})}{\sum_{h}p(z_{ih}=1)p(\mathbf{x}_i|z_{ih}=1,\widetilde{\mathbf{Z}}_{-i})} \\[1ex] & 
    \propto p(z_{ik}=1)p(\mathbf{x}_i|z_{ik}=1,\widetilde{\mathbf{Z}}_{-i}) \\[1ex]& =
    \theta_k \prod_{j\neq i}^{n} x_{ij}^{\Tilde{\alpha}_j} \dfrac{\Gamma(\sum_{j\neq i}^{n} \Tilde{\alpha}_j)}{\prod_{j\neq i}^{n} \Gamma (\Tilde{\alpha}_j)},
    \end{aligned}
\end{equation}

\noindent as $p(z_{ik}=1|\widetilde{\mathbf{Z}}_{-i})=p(z_{ik}=1)$ due to the working independence assumption.

The M-step involves finding the estimates for the mixing proportions $\boldsymbol{\theta}=(\theta_1,...,\theta_{K})$ and the Dirichlet connectivity matrix $\mathbf{A}$. The closed form solution for the mixing proportions is derived similarly to that in the standard EM algorithm by maximising $\mathbb{E}[l_c^{hyb}(\mathbf{A},\boldsymbol{\theta})]$, subject to constraint $\sum_{k} \theta_k = 1$. Let

\begin{equation}
    f = \mathbb{E}[l_c^{hyb}(\mathbf{A},\boldsymbol{\theta})] - \lambda (\sum_{k=1}^{K} \theta_k - 1).
\end{equation}

The partial derivative of $f$ with respect to $\theta_k$ is

\begin{equation}
    \dfrac{1}{\theta_k}\sum_{i=1}^{n} \hat{z}_{ik} -\lambda = 0,
\end{equation}

\noindent which gives $\lambda \theta_k = \sum_{i=1}^{n} \hat{z}_{ik}$. Summing over k and using the unit-sum constraint for $(\theta_1,...,\theta_K)$, we find that $\lambda=n$, producing the update for the mixing proportions from Equation \eqref{m-step}. 

The updates for the Dirichlet concentration parameter matrix $\mathbf{A}$ are not available in closed form and are found numerically using the R function $optim$ \citep{r_reference}. L-BFGS-B optimisation procedure \citep{Byrd_optim} is used to update the set of parameters $\{\alpha_{kh}\}_{k,h=1}^{K}$ as it allows a permitted range of values to be set; in the case of DirSBM, we are only interested in strictly positive solutions. 

\section{Integrated completed likelihood} \label{ICL}

Following \cite{ICL}, the exact complete-data integrated log-likelihood is given by

\begin{equation} \label{exact_ICL}
    \log p(\mathbf{X},\mathbf{Z}|K) = \log p(\mathbf{X}|\mathbf{Z},K) + \log p(\mathbf{Z}|K).
\end{equation}

Following closely the derivations of the ICL for random graphs in \cite{Daudin_mm_random_graphs}, we can find the first term of Equation \eqref{exact_ICL} using a BIC approximation:

\begin{equation}
    \log p(\mathbf{X}|\mathbf{Z},K) = \max_{\mathbf{A}} \log p(\mathbf{X}|\mathbf{Z},\mathbf{A},K) -\frac{1}{2} K^2 \log n(n-1),
\end{equation}
where $K^2$ is the number of parameters in the model (elements of non-symmetric $\mathbf{A}$ matrix), and $n(n-1)$ is the number of edge weights in the network (with no self-loops).

The second term of Equation \eqref{exact_ICL} is derived in the same fashion as in \cite{Daudin_mm_random_graphs}, using a non-informative Jeffreys prior and Stirling formula for an approximation of a gamma function, leading to

\begin{equation}
    \log p(\mathbf{Z}|K) = \max_{\boldsymbol{\theta}} \log p(\mathbf{Z}|\boldsymbol{\theta},K) - \frac{1}{2} (K-1) \log n,
\end{equation}
with $(K-1)$ being the number of free parameters in $\boldsymbol{\theta}$ and $n$ being the number of latent variables in the model.

Substituting the approximation results back and then replacing the complete data log-likelihood with its hybrid counterpart in the fashion of \cite{HybridML}, we arrive at 

\begin{equation}
\begin{aligned}
    ICL(K) & = \log p(\mathbf{X}|\mathbf{\hat{Z}},\mathbf{\hat{A}},K) - \frac{1}{2} K^2\log n(n-1) + \log p (\mathbf{\hat{Z}}|\boldsymbol{\hat{\theta}},K) -\frac{1}{2}(K-1)\log n \\ 
    & = \log p(\mathbf{X},\mathbf{\hat{Z}}|\mathbf{\hat{A}},\boldsymbol{\hat{\theta}},K) - \frac{1}{2} K^2\log n(n-1) -\frac{1}{2}(K-1)\log n \\
    & = l^{hyb}_c(\mathbf{\hat{A}},\boldsymbol{\hat{\theta}}|\mathbf{X},\mathbf{\hat{Z}},K) - \frac{1}{2} K^2\log n(n-1) -\frac{1}{2}(K-1)\log n.
\end{aligned}
\end{equation}

\section{Simulation studies: additional details} \label{sim_study_notes}
\begin{sloppypar}
Parameters used to generate data sets with different levels of parameter homogeneity, low, medium and high, denoted with subscripts $L$, $M$ and $H$, respectively:\medskip
\end{sloppypar}

\noindent $K=2$
\begin{equation}
    \mathbf{A}_L= \begin{pmatrix} 1.0 & 0.6 \\ 0.8 & 1.5 \end{pmatrix}, 
    \mathbf{A}_M= \begin{pmatrix} 1.0 & 0.6 \\ 0.9 & 1.4 \end{pmatrix}, 
    \mathbf{A}_H= \begin{pmatrix} 1.0 & 0.8 \\ 0.9 & 1.5 \end{pmatrix};
\end{equation}

\noindent $K=3$
\begin{equation}
    \mathbf{A}_L= \begin{pmatrix} 1.0 & 0.6 & 0.2 \\ 0.6 & 1.5 & 0.5 \\ 0.3 & 0.4 & 1.2  \end{pmatrix}, \mathbf{A}_M= \begin{pmatrix} 1.0 & 0.7 & 0.5 \\ 0.9 & 1.5 & 0.6 \\ 0.4 & 0.5 & 1.2  \end{pmatrix},     \mathbf{A}_H= \begin{pmatrix} 1.0 & 0.7 & 0.5 \\ 0.9 & 1.3 & 0.7 \\ 0.6 & 0.5 & 1.2  \end{pmatrix};
\end{equation}

\noindent $K=5$
\begin{equation}
     \mathbf{A}_L= \begin{pmatrix} 1.0 & 0.6 & 0.2 & 0.3 & 0.5 \\ 0.6 & 1.5 & 0.5 & 0.4 & 0.7 \\ 0.3 & 0.4 & 1.2 & 0.5 & 0.2 \\ 0.7 & 0.5 & 0.3 & 1.4 & 0.4 \\ 0.5 & 0.7 & 0.8 & 0.6 & 1.7  \end{pmatrix},
    \mathbf{A}_M= \begin{pmatrix} 1.0 & 0.7 & 0.5 & 0.4 & 0.6 \\ 0.9 & 1.5 & 0.6 & 0.5 & 0.7 \\ 0.4 & 0.5 & 1.2 & 0.6 & 0.3 \\ 0.8 & 0.6 & 0.4 & 1.4 & 0.5 \\ 0.5 & 0.8 & 0.9 & 0.7 & 1.7  \end{pmatrix},
\end{equation}

\begin{equation}    
    \mathbf{A}_H= \begin{pmatrix} 1.0 & 0.7 & 0.5 & 0.4 & 0.6 \\ 0.9 & 1.3 & 0.7 & 0.5 & 0.8 \\ 0.6 & 0.7 & 1.2 & 0.8 & 0.5 \\ 0.8 & 0.6 & 0.4 & 1.4 & 0.7 \\ 0.7 & 0.8 & 0.9 & 0.6 & 1.6  \end{pmatrix}.
\end{equation}

The parameter values are constrained to be positive, but some care is needed when the parameter matrices for synthetic data sets are chosen if the intention is to use such data for model testing. Due to specification of the model based on the Dirichlet distribution whose dimensions grow with the size of the network, they cannot be too large, especially in the case of large number of nodes. The expression for the complete data hybrid log-likelihood involves a gamma function term evaluated at the sum of all parameter values, and large parameters cause it to become too large to compute. The parameters also cannot be too close to 0 as we are required to evaluate the gamma function at each individual parameter value.

\section{Erasmus programme data: parameter estimates} \label{estimates_erasmus}
\subsection{K=3}

The estimate of the Dirichlet parameter matrix $\mathbf{A}$ of the 3-cluster DirSBM on the Erasmus programme data is

\begin{equation}
    \mathbf{\hat{A}}=
    \begin{pmatrix}
7.019 & 1.893 & 0.410 \\
5.325 & 2.521 & 0.535 \\
1.347 & 0.410 & 0.142 \\
    \end{pmatrix}.
\end{equation}

The expected total cluster-to-cluster exchange share and the expected node-to-node exchange share matrices are (multiplied by 100 for convenience): 

\begin{equation}
\mathbf{\hat{V}}=
    \begin{pmatrix}
50.3 & 37.3 & 12.5 \\
43.7 & 41.4 & 14.9 \\
49.8 & 33.4 & 16.8 
    \end{pmatrix},
    \hspace{3mm}
    \mathbf{\hat{W}}=
    \begin{pmatrix}
12.6 & 3.4 & 0.7 \\
8.7 & 4.1 & 0.9 \\
10.0 & 3.0 & 1.1 \\
\end{pmatrix}.
\end{equation}

The rows of these matrices correspond to the sending clusters, and the receiving clusters are along the columns. To give an example in the context of the Erasmus programme network, the entry $\hat{v}_{12} = 37.3$ indicates that 37.3\% of students of cluster 1 collectively (which contains Germany, Spain, France, Italy and the UK) go on exchange to cluster 2 collectively (which includes countries like Austria and Belgium). The entry $\hat{w}_{31}=10.0$ can be read as 10.0\% of Irish students (cluster 3) are expected to go to Germany (cluster 1), and the same percentage is expected to go from Norway (cluster 3) to Spain (cluster 1).

\subsection{K=5}

Similarly, for the $K=5$ solution, 

\begin{equation}
    \mathbf{\hat{A}}=
    \begin{pmatrix}
11.004 & 2.214 & 0.323 & 1.189 & 0.213 \\
12.197 & 4.057 & 0.847 & 1.984 & 0.282 \\
5.539 & 3.083 & 1.174 & 0.959 & 0.301 \\
3.035 & 1.298 & 0.231 & 0.158 & 0.105 \\
0.630 & 0.323 & 0.124 & 0.155&  0.108 \\
    \end{pmatrix}
\end{equation}

\noindent and

\begin{equation}
\mathbf{\hat{V}}=
    \begin{pmatrix}
39.1 & 43.3 & 5.2 & 10.6 & 1.9 \\
38.1 & 42.2 & 7.9 & 10.3 & 1.5 \\
25.1 & 51.2 & 14.2 & 7.2 & 2.3 \\
34.2 & 53.6 & 7.8 & 2.4 & 2.0 \\
24.3 & 45.8 & 14.4 & 10.0 & 5.6 \\
    \end{pmatrix},
    \hspace{3mm}
    \mathbf{\hat{W}}=
    \begin{pmatrix}
19.6 & 3.9 & 0.6 & 2.1 & 0.4 \\
12.7 & 4.2 & 0.9 & 2.1 & 0.3 \\
8.4 & 4.7 & 1.8 & 1.4 & 0.5 \\
11.4 & 4.9 & 0.9 & 0.6 & 0.4 \\
8.1 & 4.2 & 1.6 & 2.0 & 1.4 \\
\end{pmatrix}.
\end{equation}

\section{Bike sharing data: parameter estimates} \label{estimates_bike}

The Dirichlet parameter matrix $\mathbf{A}$ estimate of the 4-cluster DirSBM on the bike sharing data is

\begin{equation}
\mathbf{\hat{A}}=
    \begin{pmatrix}
1.58 & 0.25 & 0.71 & 2.04 \\
0.22 & 3.96 & 0.79 & 0.38 \\
0.61 & 1.51 & 2.16 & 1.26 \\
1.23 & 0.64 & 1.67 & 2.33 \\
    \end{pmatrix},
\end{equation}

\noindent and the expected cluster-to-cluster exchange shares and station-to-station exchange shares matrices are

\begin{equation}
\mathbf{\hat{V}}=
    \begin{pmatrix}
36.5 & 3.7 & 18.7 & 41.1 \\
5.9 & 62.3 & 23.2 & 8.6 \\
12.6 & 19.4 & 46.4 & 21.6 \\
23.2 & 7.6 & 34.3 & 34.9 \\
    \end{pmatrix},
    \hspace{3mm}
    \mathbf{\hat{W}}=
    \begin{pmatrix}
1.6 & 0.2 & 0.7 & 2.1 \\
0.2 & 4.5 & 0.9 & 0.4 \\
0.5 & 1.3 & 1.9 & 1.1 \\
1.0 & 0.5 & 1.3 & 1.8 \\
\end{pmatrix}.
\end{equation}

\bibliography{refs}

\end{document}